\RequirePackage{lineno}

\documentclass[12pt]{iopart}

\usepackage{iopams}  
\usepackage{setstack}
\usepackage{bm,bbm}
\usepackage{color}
\usepackage{amssymb}
\usepackage{verbatim}

\usepackage{setspace}

\usepackage[normalem]{ulem}

\usepackage{graphicx}  %
\usepackage{fancybox,pstricks,pst-node,pst-plot}

\newcommand{\fixsubfiglabel}[1]{(#1)}

\newcommand{\Sum}[2]{\overset{#2}{\underset{#1}{\sum}}}
\newcommand{\Int}[2]{\overset{#2}{\underset{#1}{\int}}}

\renewcommand{\d}{{\,\mathrm d}}

\newcommand{\gerdsbinomial}[2]{{{#1} \choose {#2}}}

\begin{document}
\title{Minkowski Tensors of Anisotropic Spatial Structure}
\author{
G.E.~Schr\"oder-Turk$^{1,*}$, W.~Mickel$^{1,2}$, S.C.~Kapfer$^1$, F.M.~Schaller$^1$, B.~Breidenbach$^1$, D.~Hug$^2$, and K.~Mecke$^1$
}

\address{ 
$^1$ Theoretische Physik, Friedrich-Alexander-Universit\"at Erlangen-N\"urnberg, Staudtstr.~7B, D-91058 Erlangen, Germany\\
$^2$ Karlsruhe Institute of Technology, Institute for Stochastics, Kaiserstr. 89-93, D-76128 Karlsruhe, Germany
}

\begin{abstract}
This article describes the theoretical foundation  of and explicit algorithms for  
a novel approach to morphology and anisotropy analysis of complex
spatial structure using tensor-valued Minkowski functionals, the so-called {\em Minkowski
tensors}. Minkowski tensors are generalisations of the well-known scalar
Minkowski functionals and are explicitly sensitive to anisotropic aspects of
morphology, relevant for example for elastic moduli or permeability of microstructured materials. Here we derive explicit linear-time algorithms to compute these tensorial measures for three-dimensional shapes. These apply to representations of any object that can be represented by a triangulation of its bounding surface; their application is illustrated for the polyhedral Voronoi cellular complexes of jammed sphere configurations, and for triangulations of a biopolymer fibre network obtained by confocal microscopy. The article further bridges the substantial notational and conceptual gap between the different but equivalent approaches to scalar or tensorial Minkowski functionals in mathematics and in physics, hence making the mathematical measure theoretic method more readily accessible for future application in the physical sciences.
\end{abstract}

\begin{figure}[t]
\begin{minipage}[t]{0.24\textwidth}
\includegraphics[height=0.99\textwidth]{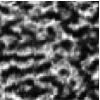}\\[0.2\baselineskip]
    \mbox{\small(a) Copolymer Film}
\end{minipage}
\nolinebreak\hspace*{0.01\textwidth}
\begin{minipage}[t]{0.24\textwidth}
    \includegraphics[height=0.99\textwidth]{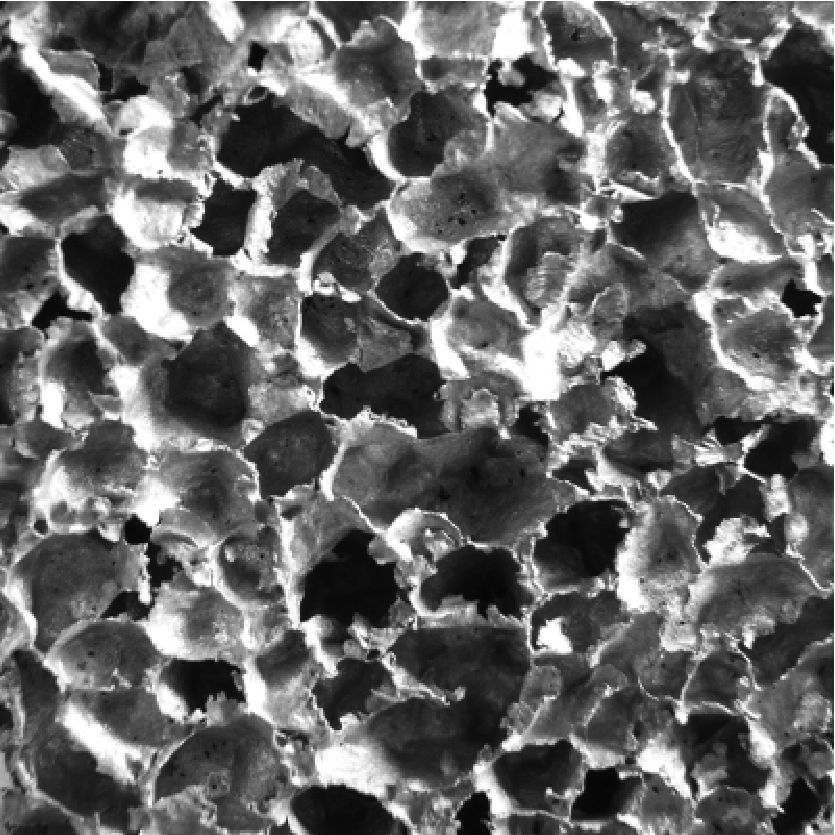}\\[0.2\baselineskip]
    \mbox{\small (b) Metal Foam}
\end{minipage}
\nolinebreak\hspace*{0.01\textwidth}
\begin{minipage}[t]{0.24\textwidth}
\includegraphics[height=0.99\textwidth]{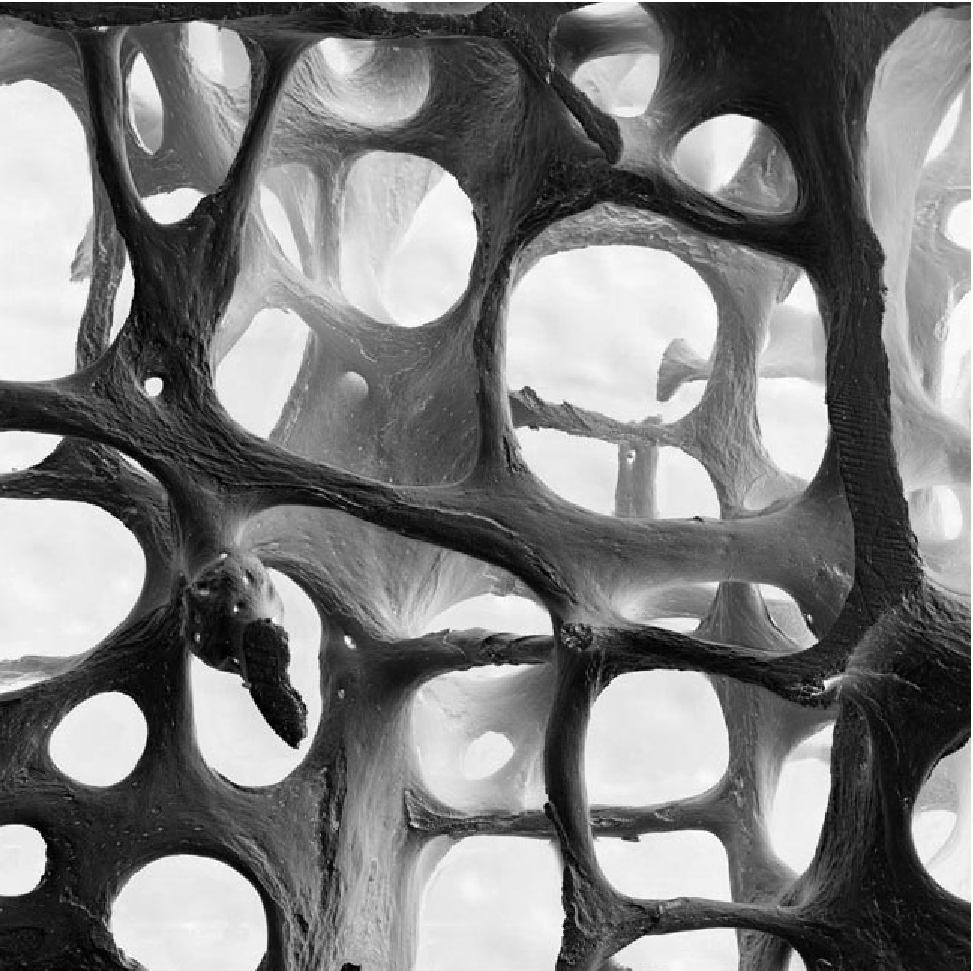}\\[0.2\baselineskip]
    \mbox{\small(c) Trabecular Bone}
\end{minipage}
\nolinebreak\hspace*{0.01\textwidth}
\begin{minipage}[t]{0.24\textwidth}
\includegraphics[height=0.99\textwidth]{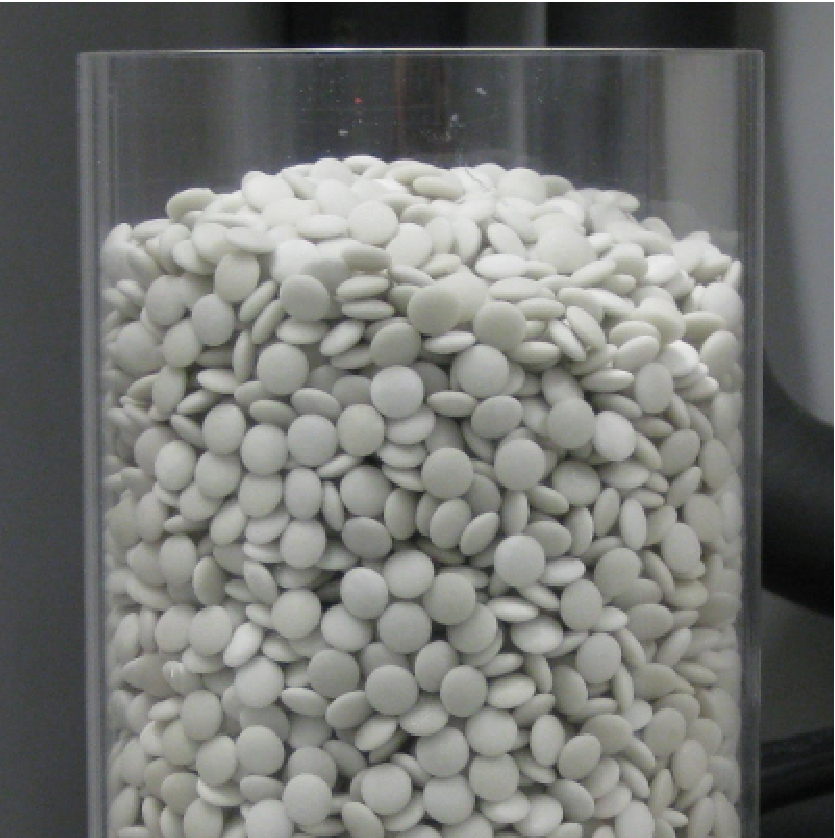}\\[0.2\baselineskip]
    \mbox{\small (d) Granular Material}
\end{minipage}
\caption{Examples of systems with anisotropic spatial structure. \fixsubfiglabel{a} A
microphase-separated copolymer film aligns under the influence of an external
electric field (image courtesy A.~B\"oker and V.~Oszowka, see also \cite{OlszowkaHundKuntermannScherdelTsarkovaBoekerKrausch:2006,OlszowkaHundKuntermannScherdelTsarkovaBoeker:2009}). \fixsubfiglabel{b} Closed-cell metal foam (image courtesy of
M.~Saadatfar
\cite{SaadatfarGarciaMorenoHutzlerSheppardKnackstedtBanhartWeaire:2009}). \fixsubfiglabel{c}
Structure of trabecular bone (image courtesy Alan Boyde \cite{BoydeEmail}). \fixsubfiglabel{d} Packing of
ellipsoids as a model system for anisotropic granular matter.}
\label{fig:examples-of-anisotropic-3D-structure}
\end{figure}

The morphology of complex spatial microstructures is often classified
qualitatively into types such as cellular, porous, network-like, fibrous,
percolating, periodic, lamellar, hexagonal, disordered, fractal, etc. Various
quantitative measures of morphology have been defined often applicable to one
specific type only, for example moments of the distributions of angles of
tangent vectors with a fixed specified direction as anisotropy characterisation
of a network structure. Apart from the concept of correlation functions, few
measures are defined sensibly and robustly for all types. In this article, we
describe the class of Minkowski tensors (MT) that apply generically to almost any type of structure which contains two or more phases separated by a well-defined interface, for example, porous media, foams, trabecular bone, granular material.
The MT are defined as integrals of powers
of normal and position vectors and surface curvatures, or curvature measures.
Because of their tensorial nature they are explicitly
sensitive to anisotropic and orientational aspects of spatial structure. Figure
\ref{fig:examples-of-anisotropic-3D-structure} shows examples of systems where
subtle anisotropy of the spatial structure influences the physical properties
and to which the analysis of this article is applicable. 

The scope of this article is the thorough theoretical description of the MT approach to spatial structure analysis and the derivation of a robust algorithm to compute MT of bi-phasic materials. It will facilitate the use of tensorial Minkowski functionals as robust structure metrics for shape description and for structure-property correlations in physics and material science. It simultaneously provides the theoretic and algorithmic basis of our previous applications of this method~\cite{SchroederTurkMickelSchroeterDelaneySaadatfarSendenMeckeAste:2010,KapferMickelSchallerSpannerGollNogawaItoMeckeSchroederTurk:2010,SchroederTurkMickelVarslotDiCampoFogdenHyde:2011,SchroederTurk:2011b, MickelSchroederTurkMecke:2012, SchroederTurkKapferBreidenbachBeisbartMecke:2009,KapferMickelMeckeSchroederTurk:2012}, and broadens the scope of the MT concept. A secondary purpose is to bridge the gap between the notation and concepts commonly used in the physics literature for scalar Minkowski functionals, based on surface and volume integrals, and the integral geometry literature, where both scalar and tensorial Minkowski functionals are derived based on measure theory. \\

Minkowski tensors are direct generalisations of scalar-valued Minkowski functionals (MF). These latter are well established as succinct descriptors of
morphology and spatial structure for various physical processes
\cite{Mecke:2000}. These integral geometric measures have been applied to
disordered porous materials \cite{ArnsKnackstedtPinczewskiMecke:2004} and are
relevant to flow phenomena therein \cite{MeckeArns:2005,ScholzWirnerGoetzRuedeSchroederTurkMeckeBechinger:2012}, to nano-scale
microstructures in copolymers \cite{RehseMeckeMagerle:2008}, to the dewetting
dynamics of thin films \cite{BeckerGruenSeemannMantzJacobsMeckeBlossey:2003},
and to Turing patterns \cite{Mecke:1996}. They have also been shown to be the
most pertinent morphological quantities on which the thermodynamic properties of
simple fluids near curved solid interfaces depend
\cite{KoenigRothMecke:2004,KoenigBrykMeckeRoth:2005}. The mathematical theory of
MF and their generalisations has been comprehensively
developed in the context of integral geometry
\cite{Hadwiger:1957,Matheron:1975,Santalo:1976,SchneiderWeil:2008}, with several aspects shared also with the discipline of mathematical morphology \cite{Serra:1983,Heijmans:1994,Soille:1999}.\\

MF as scalar quantities are not sensitive to features of the
morphology which relate to orientation or directional anisotropy, since
motion-invariance is one of their defining properties. Therefore scalar MF do not
provide quantification of anisotropy that is relevant to study the
direction-dependence of physical processes, such as elastic properties or
permeability of anisotropic porous or microstructured materials or systems with
external fields. This motivates their generalisation to tensorial quantities.
MT have already been shown to be the relevant morphological
descriptors for a density functional theory of fluids of non-spherical particles
\cite{HansenGoosMecke:2009} and of DNA conformations
\cite{HansenGoosRothMeckeDietrich:2007}, and of a simple model for transport with
molecular motors \cite{SporerGollMecke:2008}. They have also been used, in 2D,
as morphology descriptors of arrangements of neuronal cells
\cite{BeisbartBarbosaWagnerCosta:2006}, galaxies
\cite{BeisbartDahlkeMeckeWagner:2002}, and Turing patterns
\cite{SchroederTurkKapferBreidenbachBeisbartMecke:2009}. 

The mathematical
discipline of integral geometry has proven statements regarding continuity and
completeness similar to the Hadwiger theorem in the scalar case
\cite{Alesker:1999,HugSchneiderSchuster:2008,Mueller:1953,McMullen:1997}.
However, an algorithm for the computation of the MT applicable to
experimental 3D data -- a prerequisite for their use as shape indices for
experimental data -- has thus far been lacking. (Note that the work in refs.~\cite{SchroederTurkMickelSchroeterDelaneySaadatfarSendenMeckeAste:2010,KapferMickelSchallerSpannerGollNogawaItoMeckeSchroederTurk:2010,SchroederTurkMickelVarslotDiCampoFogdenHyde:2011,SchroederTurk:2011b, MickelSchroederTurkMecke:2012, SchroederTurkKapferBreidenbachBeisbartMecke:2009,KapferMickelMeckeSchroederTurk:2012,MickelSchroederTurkMecke:2012} has employed the algorithms described in this article without thorough comprehensive description.) 

A primary application of rank-2 MT is the quantitative analysis of the
degree of intrinsic anisotropy of materials with complex spatial structure.
Scalar  measures of anisotropy are easily derived as eigenvalue ratios of the MT.
Evidently, alternative methods for the characterisation of anisotropy and
alignment exist. Fourier transforms are a common way to characterise anisotropy,
and have been applied e.g.~for trabecular bone
\cite{BrunetImbaultLemineurChappardHarbaBenhamou:2005}, for electrodeposited
patterns \cite{SaitouFukuoka:2005}, for fibre systems \cite{TunakLinka:2007},
and for structured polyethylene mats \cite{BlackadderGrayMcCrum:1976}. Related
methods based on correlation functions are also known
\cite{AtanasoDurrellVulkovaBarberYordanov:2006,Berryman:1998}. Fourier methods that
analyse the amplitude of the Fourier transform
of a gray-scale image in polar coordinates can also quantify alignment, e.g.~of
copolymer films in electric fields
\cite{OlszowkaHundKuntermannScherdelTsarkovaBoekerKrausch:2006}. Anisotropy indices derived from the normal vector
distribution of a given shape, similar to the MT, have been used
to describe the shape anisotropy of simulated 3D foam cells
\cite{KraynikReineltVanSwol:2003, evans2012} and liquid interfaces \cite{DoiOhta:1991, SchroederTurkMickelVarslotDiCampoFogdenHyde:2011}. An
anisotropy measure applicable to porous media is derived from the directional
variations of average chord lengths. For a binary composite, i.e.~consisting of
a solid and a void phase, a chord is a segment of an infinite straight line that
is fully contained in one of the two phases. Analyses of chord lengths and the
derived mean intercept length ellipsoid are used for the  investigation of the
microstructure of bone
\cite{Whitehouse:1974,HarriganMann:1984,WaldVasilicSahaWehrli:2007,
MathieuMuellerBourbanPiolettiMullerManson:2006,
ChiangWangLandisDunkersSnyder:2006,InglisPietruszczak:2003}, see also
ref.~\cite{KetchamRyan:2004} for a comparison of anisotropy measures based on
mean-intercept length, star-volume and star-length distributions. Deformations
of cellular or granular material have recently been quantified using the
so-called {\em texture tensor} $C$, defined as the sum $C_{ij}:=\sum l_i l_j$ over a subset of link vectors $\mathbf{l}$ in the structure
\cite{AubouyJiangGlazierGraner:2003,GranerDolletRaufasteMarmottant:2008}. The
texture tensor can be used to characterise anisotropy, e.g.~for Antarctic ice
crystals \cite{DurandIceCore:2004} and liquid foam cells
\cite{JaniaudGraner:2005}. Further anisotropy measures are based on the Steiner
compact \cite{RatajSaxl:1989}, wavelet analysis \cite{SungFarnood:2007}, the
orientation of volumes \cite{OdgaardJensenGundersen:1990} or star-volumes
\cite{KarlssonCruzOrive:2005}. Two-dimensional equivalents of the anisotropy
measures discussed in this article have previously been used for the analysis of
the shape of neuronal cells \cite{BeisbartBarbosaWagnerCosta:2006} and galaxies
\cite{BeisbartDahlkeMeckeWagner:2002}, and are discussed in detail in
\cite{SchroederTurkKapferBreidenbachBeisbartMecke:2009}.

The paper is organised as follows: Section \ref{sec:def-and-basic-props}
provides an overview of the theory of MF and MT, based
on their definition by surface integrals which is widely used in the physical
sciences. To bridge the gap in notation between the physics and the mathematics
literature this section also includes a discussion of the alternative definition
based on measure theory. Section \ref{sec:algo-for-triangul-bodies} derives
algorithms to compute MT for bodies represented by triangulations of their bounding surface; their implementation is provided as supplementary online material to this article. Section
\ref{sec:anisotropy-measures} describes anisotropy measures derived from the
MT and illustrates their application to two experimental data
sets. The appendix provides analytic expressions for the MT for
some simple geometric shapes.

\section{Definition and fundamental properties}
\label{sec:def-and-basic-props}

Scalar MF and MT can be used as shape measures to quantify the shape or form of an object.
They can be defined in two largely equivalent ways. In the physical sciences, a
definition based on curvature-weighted integrals of position or surface normal
vectors over the object's bounding surface has been popular for the scalar MF, and forms the basis
of the numerical algorithms derived in this article. An alternative, more
fundamental definition is provided by the measure theoretic approach of integral
geometry (see \ref{subsec:def-of-MT-fundam-measure}).

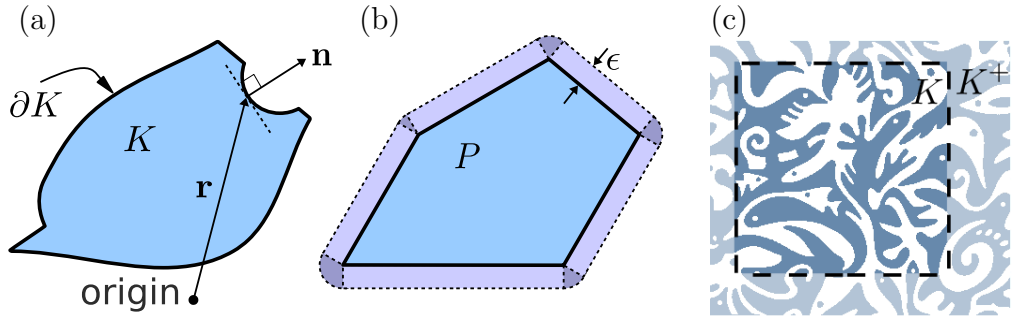
\begin{figure}[t]
\setlength{\unitlength}{0.85\textwidth}
\hfill
\begin{picture}(1,0.41)(0,0)
\put(0,0.07){
\includegraphics[width=0.8\unitlength]{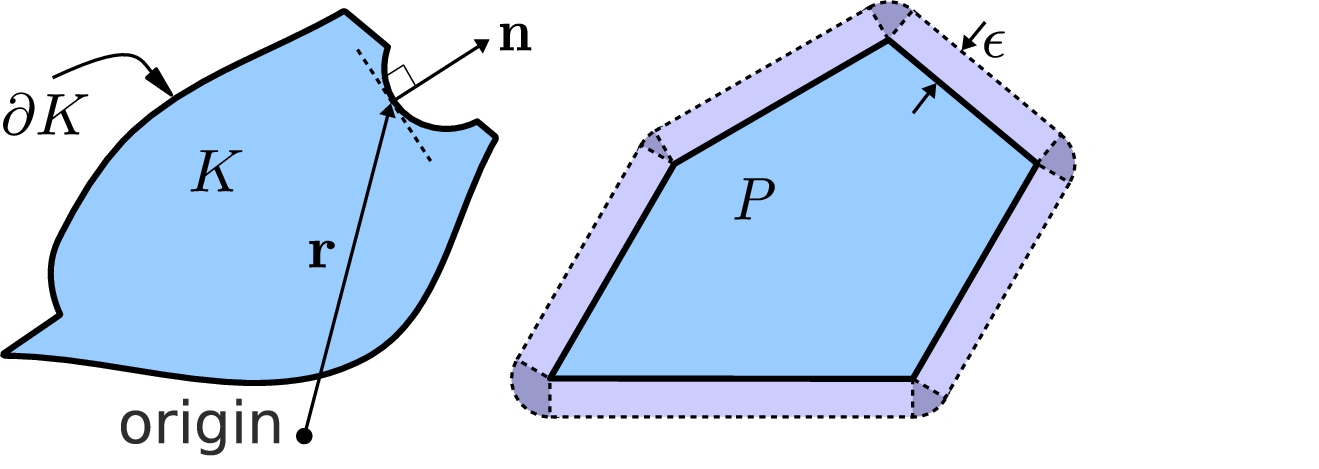}
} %
\put(0.7,0.07){
\includegraphics[width=0.3\unitlength]{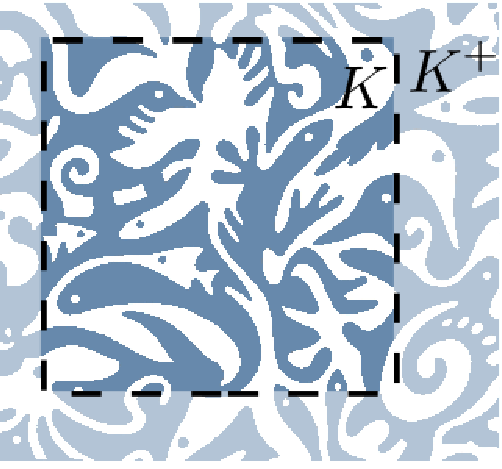}}
\put(0.02,0.36){\fixsubfiglabel{a}}
\put(0.36,0.36){\fixsubfiglabel{b}}
\put(0.71,0.36){\fixsubfiglabel{c}}
\end{picture}
\caption{\fixsubfiglabel{a} A body $K$ with bounding surface $\partial K$; \fixsubfiglabel{b} a convex
polytope $P$ and its parallel body (or dilation) $P\uplus B_\epsilon$; \fixsubfiglabel{c} a subset of a
topologically more complex body based on Craig Marlow's painting ``White
Spirits'' \cite{Marlow:2004}. The latter demonstrates a common experimental
situation, namely that the given body $K$ represents a finite subset of a larger
body $K^+$, here  the body $K$ clipped to the {\em window of
observation} $T$, $K=K^+\cap T$. $K$ is assumed to be a
representative subset of the larger body, which allows for the estimation of
intrinsic shape features of $K^+$. If only $K$, but not
$K^+$, is available for analysis particular care must be taken w.r.t.~those
bounding surface patches of $K$ that result purely from taking the subset,
i.e.~those that are on the boundaries of the window of observation. }
\label{Fig:Body-illu}
\end{figure}

The object, also referred to as {\em body}, whose shape can be characterised by
MT or MF is denoted by $K$. Assuming that $K$ is a compact set with nonempty interior embedded in Euclidean
space $\mathbbm{R}^3$ and is bounded by a sufficiently smooth
surface $\partial K$, we define MF of
$K$ as
\begin{equation}
W_0(K) = \Int{K}{} \d V \quad \mbox{and} \quad W_\nu(K) =
\frac{1}{3}{\Int{\partial K}{} G_{\nu} \d A}
\label{Eqn:Def-ScalarMeasures}
\end{equation}
in space-dimension $d=3$ and with $\nu=1,2,3$. The scalar functions $G_\nu$ are
$G_1=1$, the mean curvature $G_2=(k_1+k_2)/2$ and the Gaussian
curvature $G_3=k_1\cdot k_2$ of the bounding surface $\partial K$; the scalars $k_1,k_2$ are the principal curvatures on $\partial K$ as defined in
differential geometry, $\d V$ is the infinitesimal volume and $\d A$ the scalar
infinitesimal area element. This definition naturally applies to both convex and
non-convex bodies of arbitrary topology with a sufficiently smooth bounding
surface. The prefactor is chosen such that for a sphere $B_R$ with
radius $R$ the scalar MF are
$W_\nu(B_R)=\kappa_3\,R^{3-\nu}$ where $\kappa_3=4\pi/3$ is the volume of
the 3-dimensional unit sphere \footnote{Other normalisations of the scalar MF are also common in the literature.
The kinematic formulae
\cite{Santalo:1976,SchneiderWeil:2008}
have particularly simple coefficients if the normalisation
$M_\nu(K)=\kappa_{d-\nu} W_\nu(K) / (\kappa_\nu \kappa_d)$ is
used, with the volume of the $n$-dimensional unit ball $\kappa_n:=\pi^{n/2} /
\Gamma(n/2+1)$.
In the mathematical literature, in $d$-dimensional Euclidean space the
normalisation
$V_{d-\nu}(K) =  {d \choose \nu} W_{\nu}(K) /\kappa_{\nu}$ is frequently used,
and the $V_{d-\nu}$ are called the {\it intrinsic volumes} of $K$.  In three
dimensions,
the set of MF thus consists of the volume $W_0=M_0=V_3$,
the surface area $3W_1=8M_1=2V_2$, the integrated mean curvature
$3W_2=2\pi^2M_2=\pi
V_1$, and the Euler characteristic $\chi = {3\over 4\pi} W_3={4\pi\over
3}M_3=V_0$ with $\chi(B_R)=1$.}.

The MF $W_2$ and $W_3$ (in 3D space) are not properly defined by
eq.~(\ref{Eqn:Def-ScalarMeasures}) for bodies with sharp edges or corners, due
to singularities of the mean and Gaussian curvatures $G_2$ and $G_3$. However,
for convex bodies, consideration of a parallel or dilated body in the limit of
vanishing thickness provides a robust definition, by use of the Steiner formula.
The Steiner formula states that, for a given convex body $K$, the MF of the
parallel or dilated body $(K\uplus B_\epsilon)$ are a polynomial in $\epsilon\ge 0$ with coefficients proportional to the MF of $K$; for $\epsilon > 0$, 
$K_\epsilon := K\uplus B_\epsilon := \lbrace
\mathbf{x}_1+\mathbf{x}_2 : \mathbf{x}_1\in K,\mathbf{x}_2\in B_\epsilon
\rbrace$ is the parallel or dilated body of $K$ (see Fig.~\ref{Fig:Body-illu}).
Specifically for the volume one finds $W_0(K\uplus B_\epsilon)=W_0(K)+
3W_1(K)\epsilon+ 3W_2(K)\epsilon^2+W_3(K)\epsilon^3$, and more
generally
\begin{equation}
W_\nu(K\uplus B_\epsilon)=\sum_{\mu=\nu}^{3} {3-\nu \choose \mu-\nu}  W_\mu(K)
\epsilon^{\mu-\nu}.
\label{Eqn:SteinersFormula}
\end{equation}
 Sharp edges and vertices of $K$ correspond to
cylindrical or spherical segments on the bounding surface $\partial(K\uplus
B_\epsilon)$ of the parallel or dilated body. The bounding surface is sufficiently smooth
for $\epsilon>0$ and  $W_\nu(K\uplus B_\epsilon)$ converges to $W_\nu(K)$ in the limit $\epsilon\rightarrow0$. It is
further necessary to define MF and MT for certain non-convex bodies, with or without
positive reach, see e.g.~\cite[Note 1 to Section 5.3 and
refs.~therein]{SchneiderWeil:2008}; this is achieved below by exploiting an
additivity relationship. A further discussion for non-smooth bodies can be found in
section \ref{sec:algo-for-triangul-bodies}.

MT are symmetric tensors (that is, invariant under index permutation), which are generated by
symmetric tensor products of
position vectors $\mathbf{x}$ and normal vectors $\mathbf{n}$ of $\partial K$. 
The dyadic (or tensor) product of two vectors $\mathbf{a}$ and $\mathbf{b}$ is
$(\mathbf{a}\otimes\mathbf{b})_{ij}:=\mathbf{a}_i\mathbf{b}_j$.
Let now $\mathbf{a}$ and
$\mathbf{b}$ be symmetric tensors of rank $r$ and $s$, respectively. The symmetric tensor product is defined as
\begin{equation}
\left(\mathbf{a}\odot \mathbf{b} \right)_{i_{1}\ldots i_{r+s}}:= \frac{1}{\left(r+s\right)!}\sum_{\sigma\in S_{r+s}} \mathbf{a}_{i_{\sigma(1)}}\cdots \mathbf{a}_{i_{\sigma(r)}} \mathbf{b}_{i_{\sigma(r+1)}}\cdots \mathbf{b}_{i_{\sigma(r+s)}},
\end{equation}
where $S_{r+s}$ is the permutation group of $r+s$ elements. For two
tensors $\mathbf{a}$ and $\mathbf{b}$, we use the shorter notation
$\mathbf{a}^2:=\mathbf{a}\odot\mathbf{a}=\mathbf{a}\otimes\mathbf{a}$ and
$\mathbf{ab}:=\mathbf{a}\odot\mathbf{b}$. For example, if $\mathbf{a}$ and
$\mathbf{b}$ are both vectors, the symmetric tensor product is the tensor $\mathbf{ab}$ of rank 2.

The MT of rank two are defined as
\begin{eqnarray}
W_0^{2,0}(K) &:=& \Int{K}{} \mathbf{x}^2 \, \d V \label{Eqn:DefVolumeTensors} \\
W_\nu^{r,s}(K) &:=& \frac{1}{3}\Int{\partial K}{} G_\nu\,
\mathbf{x}^r\mathbf{n}^s \, \d A.\label{Eqn:DefSurfaceTensors}
\end{eqnarray}
with $\nu=1,2,3$ and 
$(r,s)=(2,0)$, $(1,1)$ or $(0,2)$.
For ease of notation, we set $W_0^{r,s}:=0$ for $s>0$ and $W_\nu^{r,s}:=0$ if
$\nu<0$ or $\nu>3$.
For a three-dimensional body, this definition yields 10 MT, not
counting the ones that vanish by definition for all bodies.

MT of rank one (called Minkowski vectors) are defined by
$W_0^{1,0}:=\int_K \mathbf{x} \d V$
and by $W_\nu^{1,0}:= \frac 13\int_{\partial K} \mathbf{x} \d A$ for
$\nu=1,2,3$.
The prefactors are chosen such that, 
for a sphere centred at $\mathbf{C}$, the so-called {\em curvature centroids}
$W_\nu^{1,0}/W_\nu$ are equal to $\mathbf{C}$. Note specifically that $W_1^{1,0}/W_0$ is the
 centre of mass (assuming a solidly filled body of constant density).
Formally, vectors $W_\nu^{0,1}$ proportional to $\int_{\partial K} \mathbf{n}  \d A$
for $\nu=1,2,3$ are also defined, however they vanish for any body with a
closed bounding surface \cite{Mueller:1953}.

MT are isometry covariant, that is their behaviour under
translations and rotations is given by
\begin{eqnarray}
W_\nu^{r,s}(K\uplus\mathbf{t})&=&\sum_{p=0}^r {r \choose p} {\mathbf{t}^p\,
W_\nu^{r-p,s}(K)}
\label{Eqn:translation-covariance}\\
W_\nu^{r,s}(U K)&=&\hat{U}_{r+s} : W_\nu^{r,s},
\label{Eqn:rotation-covariance}
\end{eqnarray}
where $K\uplus\mathbf{t}$ is the translation of $K$ by the vector $\mathbf{t}$,
$U K$ is a rotated copy of $K$,
and $\hat U_{r+s}$ denotes the corresponding rotation tensor for a
rank-$(r+s)$ tensor:
\begin{equation}
\left(\hat{U}_{r+s} : W_\nu^{r,s}\right)_{i_1,\ldots,i_{r+s}}:= \sum_{j_1,\ldots,j_{r+s}} U_{i_1,j_1}\ldots U_{i_{r+s},j_{r+s}}\left(W_\nu^{r,s}\right)_{j_1,\ldots,j_{r+s}} \,.
\label{Eqn:rotation-operator}
\end{equation} 
In this expression, $U_{ij}$ is the conventional orthogonal $3\times 3$
transformation matrix associated with the rotation $U$.

\begin{table}[t]
\begin{minipage}{\columnwidth}
\vspace{0.5em}
\end{minipage}
\renewcommand{\arraystretch}{1.4}
\begin{tabular*}{\columnwidth}{@{\extracolsep{\fill}}  l|l|l|l l }
\hline
Homogeneity \linebreak {[unit]}	& rank 0	& rank 1		& rank 2
	& translation \linebreak behaviour\\ \hline\hline
$\lambda^5$ $[m^5]$	& --		& --			& $W_{0}^{2,0}$
& genuinely translation covariant~\\ \hline
$\lambda^4$ $[m^4]$	& --		& $W_{0}^{1,0}$ 	& $W_{1}^{2,0}$
& genuinely translation covariant~\\ \hline
$\lambda^3$ $[m^3]$	& --		& $W_{1}^{1,0}$ 	& $W_{2}^{2,0}$
& genuinely translation covariant~\\ 
			& $W_0$		& --		 	& $W_0\,Q$
& translation invariant~\\ \hline
$\lambda^2$ $[m^2]$	& --		& $W_{2}^{1,0}$ 	& $W_{3}^{2,0}$
& genuinely translation covariant~\\  
			& --		& --		 	& $W_{1}^{0,2}$
& translation invariant~\\
			& $W_1$		& --		 	& $W_1\,Q$
& translation invariant~\\ \hline
$\lambda^1$ $[m^1]$	& --		& $W_{3}^{1,0}$ 	& --
& genuinely translation covariant~\\
			& --		& --		 	& $W_{2}^{0,2}$
& translation invariant~\\ 
			& $W_2$		& --		 	& $W_2\,Q$
& translation invariant~\\ \hline
$\lambda^0$ $[1]$	& $W_3$		& --		 	& $W_3\,Q$
& translation invariant~\\ \hline
\end{tabular*}
\renewcommand{\arraystretch}{1.}
\caption{Basic tensor valuations in 3D.
The Minkowski functionals (scalars, rank 0) are motion invariant, the Minkowski vectors
(rank 1) are genuinely translation covariant. For rank two, the space
of Minkowski tensors decomposes into two complementary subspaces according
to translation behaviour (indicated by the last column): genuinely
translation covariant and translation invariant tensors.   The latter include
tensors obtained by multiplying the scalar MF $W_\nu$ with
the unit tensor $Q:=\mathbf e_1^2+\mathbf e_2^2+\mathbf e_3^2$ of rank two,
where $\mathbf e_1,\mathbf e_2,\mathbf e_3$ are vectors of an orthonormal basis of
$\mathbbm{R}^3$.}
\label{Tab:Minkowski-Tensoren-Basis}
\end{table}

For $r=0$, equation (\ref{Eqn:translation-covariance}) gives
$W_\nu^{r,s}(K)=W_\nu^{r,s}(K\uplus\mathbf{t})$. A tensor that fulfils this
relation for all $K$ is called {\em translation invariant}.
 Genuinely translation covariant tensors fulfil eq.~(\ref{Eqn:translation-covariance}) but not
the translation invariance condition.
 For the sake of
brevity, we will use the term {\em translation covariant} to denote specifically
the genuinely translation covariant tensors. All $W_\nu^{0,s}$ are translation invariant
by their definition; in dimension $d=3$, also the tensors $W_1^{1,1}, W_2^{1,1}$ and
$W_3^{1,1}$ are
translation invariant due to the envelope theorems of M\"uller
\cite{Mueller:1953}. Translation invariance is important whenever a natural choice for
the origin is not available. 

All MF and MT are homogeneous, i.e.~they fulfil the
homogeneity relation $W_\nu^{r,s}(\lambda K)=\lambda^{3+r-\nu}W_\nu^{r,s}(K)$.
Table~\ref{Tab:Minkowski-Tensoren-Basis} specifies the translation and
homogeneity behaviour of the MF and MT.

Thus far, MT have been defined for (a) convex or non-convex bodies with a smooth
bounding surface and (b) convex bodies which may have sharp
corners and edges. The case of non-convex bodies with concave sharp corners or
edges cannot be treated by the parallel body construction (``dilation'') without further
assumptions (technically, such bodies do not represent {\em sets of positive
reach} \cite{SchneiderWeil:2008, RatajZaehle:1995}). An extension of the
definition of MF and MT to finite unions of convex      
bodies is achieved by exploiting a property called {\em additivity}
\begin{equation}
W_{\nu}^{r,s}(K\cup K') = W_{\nu}^{r,s}(K)  + W_{\nu}^{r,s}(K') -
W_{\nu}^{r,s}(K\cap K'),
\label{eq:additivity}
\end{equation}
if $K$, $K'$ are all convex. In general, 
the union $(K\cup K')$ of two arbitrary convex bodies $K$ and $K'$ is not convex
while the intersection $(K\cap K')$ is convex. Since $W_\nu^{r,s}$ are continuous functionals on convex bodies, eq.~(\ref{eq:additivity}) can
be used (see Groemer's extension theorem \cite{SchneiderWeil:2008}) to define the MF and MT for
bodies that are unions of a finite number of convex bodies (such sets
are called polyconvex).

MT  of rank ($r+s$) with $r+s>1$ are \emph{not} completely linearly
independent, i.e.~they do not contain independent shape information
\cite{HugSchneiderSchuster:2008,HugSchneiderSchuster:2008b}. For rank-2 MT and $d=3$ the
linear dependencies
\begin{equation}
Q W_\nu(K)=\nu
W_\nu^{0,2}(K)+(3-\nu)W_{\nu+1}^{1,1}(K)\label{Eqn:LinearDependencyRankTwo}
\end{equation}
are valid for any polyconvex body $K$ in $\mathbbm{R}^3$ and
$\nu=0,1,2,3$; $Q$ is the unit tensor of rank 2. In particular, it follows that $Q W_3=3 W_3^{0,2}$.
Specifically, for $\nu=0$
one obtains a special case of the Gauss' theorem 
\begin{equation}\label{eq:GaussTheoremFromLinDep}
Q W_0=3 W_1^{1,1} \Longrightarrow W_0=\mathrm{tr}\,W_1^{1,1}.
\end{equation}
These relations are special cases of \cite[eq.~(1.1)]{HugSchneiderSchuster:2008}
or \cite[eq.~(1.5)]{HugSchneiderSchuster:2008b}.

Alesker's theorem \cite{Alesker:1999} makes a strong statement about the
completeness of the MT for the purpose of shape description.
For the special case of tensors of rank two, it states that any
isometry covariant, additive, continuous 
functional $\varphi$ on general convex bodies in $\mathbbm{R}^3$, taking values
in the 
space of symmetric tensors of rank two over $\mathbbm{R}^3$, is a linear
combination of the basic tensor valuations (an additive functional is called \emph{valuation}) 
$Q^pW_\nu^{r,s}$, that is
\begin{equation}\renewcommand{\arraystretch}{0.7}
    \varphi(K) = \sum_{\nu,r,s,p} \varphi_\nu^{r,s,p} Q^p W_\nu^{r,s}(K) 
\label{eq:aleskers-theorem}
\end{equation}
with real coefficients $\smash{\varphi_{\nu}^{r,s,p}}$ that do not depend on the 
convex body $K$ \cite[p.~150]{HugSchneiderSchuster:2008}.  The coefficients $\varphi_\nu^{r,s,p}$ vanish unless
$r+s+2p = 2$.  Starting from eq.~(\ref{eq:aleskers-theorem}) and using the
linear dependencies among the 
basic tensor valuations, $\varphi$ can be expressed in terms of linearly
independent 
basic tensor valuations which form a basis of the corresponding vector space.
The vector space of continuous, isometry covariant tensor valuations
of rank two in $\mathbbm{R}^3$ has dimension 10. A particular basis of this
vector space consists of the six
tensor valuations $W_0^{2,0}$, $W_1^{2,0}$, $W_2^{2,0}$, $W_3^{2,0}$,
$W_1^{0,2}$ and $W_2^{0,2}$, which contain
pertinent independent shape information, and of the four tensor valuations $Q
W_\nu$, $\nu=0,\ldots,3$. A summary is provided in
Table~\ref{Tab:Minkowski-Tensoren-Basis}.

Since $\varphi$ is continuous and additive on convex bodies, it can be extended
as an additive 
functional to finite unions of convex bodies. For this additive extension,
eq.~(\ref{eq:aleskers-theorem}) 
remains valid since the right-hand side is a linear combination of additive
functionals. 
It should be emphasised, however, that although all of these functionals are
continuous on 
the space of convex bodies, they are not continuous on the space of finite
unions of convex bodies, see the example in
\cite[Fig.~3]{SchroederTurkKapferBreidenbachBeisbartMecke:2009}.

\subsection{MT of convex polytopes}
\label{subsec:Mink-Tenso-of-conv-polyh}

\begin{figure}[t]
\begin{minipage}{0.33\textwidth}
$W_0^{2,0}$ {\footnotesize -- moment tensor solid} \\[1em]
\begin{center}\includegraphics[width=0.67\textwidth]{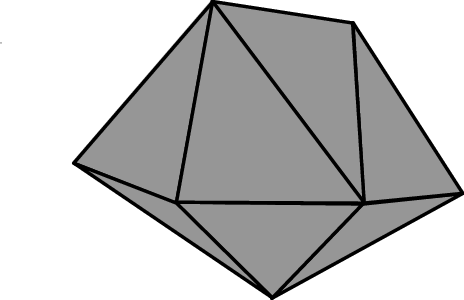}\end{center}
\end{minipage}\hfill
\begin{minipage}{0.33\textwidth}
$W_1^{2,0}$  {\footnotesize -- moment tensor hollow}\\
\begin{center}\includegraphics[width=0.67\textwidth]{./w020.eps}\end{center}
\end{minipage}
\begin{minipage}{0.33\textwidth}
$W_0^{2,0}$ {\footnotesize -- moment tensor wire-frame}\\
\begin{center}\includegraphics[width=0.67\textwidth]{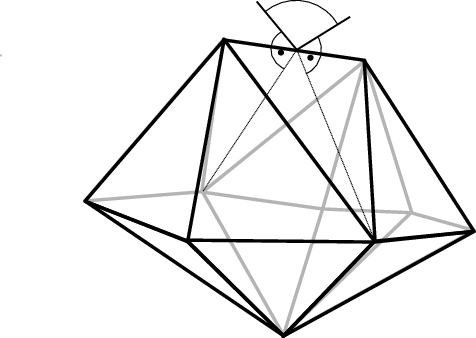}\end{center}
\end{minipage}\hfill
\begin{minipage}{0.33\textwidth}
$W_3^{2,0}$ {\footnotesize -- moment tensor vertices}\\
\begin{center}\includegraphics[width=0.6\textwidth]{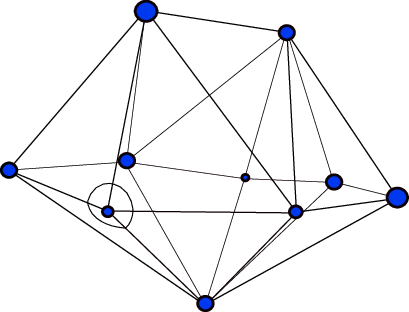}\end{center}
\end{minipage}
\begin{minipage}{0.33\textwidth}
$W_1^{0,2}$ {\footnotesize -- normal distribution}\\
\begin{center}\includegraphics[width=0.67\textwidth]{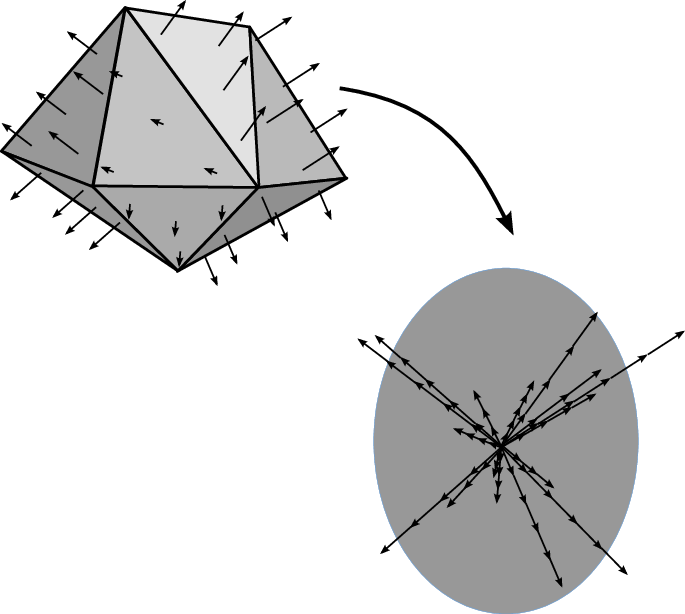}\end{center}
\end{minipage}\hfill
\begin{minipage}{0.33\textwidth}  
$W_2^{0,2}$ {\footnotesize -- curvature distribution}\\
\begin{center}\includegraphics[width=0.83\textwidth]{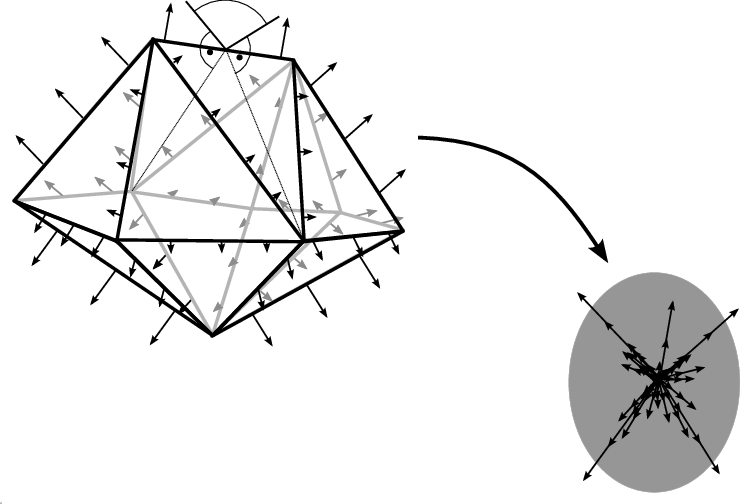}\end{center}
\end{minipage}
  \caption{Interpretations of the Minkowski tensors for a polytope $P$: \fixsubfiglabel{a} Assuming homogeneous density, $W_2^{2,0}$ is the mass distribution tensor.
\fixsubfiglabel{b-d} Contributions to surface-integrated moment tensors $W_\nu^{2,0}$ with $\nu=1,\ldots,3$
are concentrated on $(3-\nu)$-dimensional surfaces. \fixsubfiglabel{e-f} Moments of the normal distribution on $\partial P$.
Contributions to normal moment tensors $W_\nu^{0,2}$ with $\nu=1,\ldots,2$
are also concentrated on $(3-\nu)$-dimensional surfaces. (reproduced from ref.~\cite{SchroederTurk:2011b})}
  \label{fig:illu-tensors-fabian}
\end{figure}

It is instructive to illustrate the MT for convex polytopes $P$
 and to point out similarities to the tensor of inertia.
We use here the letter $P$, rather than $K$, to denote a body whose bounding
surface is a polytope, i.e.~consisting of piece-wise linear facets.
For a polytope,
the tensors $W_\nu^{2,0}$ characterise the distribution of mass if the cell is
a solid cell ($W_0^{2,0}$), a hollow cell ($W_1^{2,0}$), a wire frame
($W_2^{2,0}$) and
a cell consisting of points at the vertices only ($W_3^{2,0}$); in the last two
cases, however, this distribution of mass is weighted with certain exterior
angles (see Fig.~\ref{fig:illu-tensors-fabian}).

The tensor of inertia $\mathbbm{I}$, defined by $\mathbbm{I}_{ij}=\int_K \left(
-\mathbf{x}_i\,\mathbf{x}_j+\delta_{ij}|\mathbf{x}|^2\right) \d V$, is a measure
of the mass distribution of a (homogeneous) body $K$, relevant for the
relationship between a rotation and the resulting moment. As $\mathbbm{I}$ is
not translation-covariant, it is not a linear combination of the MT. However the simple relationship
$\mathbbm{I}(K)=-W_0^{2,0}(K)+\mathrm{tr}\left(W_0^{2,0}(K)\right)Q$ holds for
arbitrary
$K$; as above $Q$ is the unit tensor of rank two of
$\mathbbm{R}^3$, also called {\em metric tensor}. This illustrates that $W_0^{2,0}(K)$ is a measure of the
distribution of mass if $K$ is a homogeneously filled solid, somewhat analogous to the tensor of inertia. Similarly,
$W_1^{2,0}(K)$ characterises the mass distribution if $K$ is hollow and bounded
by an infinitesimally thin surface sheet. This interpretation of $W_0^{2,0}$
and $W_1^{2,0}$ is valid for bodies $K$ bounded by arbitrary surfaces (not
just polytopes).

For a polytope $P$, the tensor $W_2^{2,0}(P)$ reduces to a line integral over
the edges of the polytope (as the mean curvature vanishes on the flat facets,
see also section \ref{sec:algo-for-triangul-bodies}) and is hence related to a
mass distribution if $P$ is given by a wire frame with wires along the edges.
However, imposed by the requirement of additivity, the weight of the wire cannot be
uniform but must be proportional to the mean curvature along the edge (i.e.~the
dihedral angle). Similarly, the tensor $W_3^{2,0}(P)$ reduces to a sum of point
contributions, as the Gaussian curvature $G_3$ of $P$ vanishes except at the
vertices of the given polytope $P$. Again due to the additivity requirement, these vertices need to be weighted
with the Gaussian curvature $G_3$.

\subsection{Definition based on fundamental measure theory}
\label{subsec:def-of-MT-fundam-measure}

This section describes the alternative (and in some sense more fundamental)
definition of MF and MT based on integral geometry and fundamental measure
theory. The purpose of this section is to bridge the gap between the mathematics
and physics literature on MF and MT. However, its content is not required for
the numerical approaches to MF and MT described in
section~\ref{sec:algo-for-triangul-bodies}.

In integral geometry, the definition of MF and MT is based on so-called {\em
support measures} (sometimes called  {\em generalised curvature measures}) that can be thought of
as local versions of the scalar
MF \cite{Hadwiger:1957,Santalo:1976,SchneiderBrunnMinkowski,
SchneiderWeil:2008}.

\begin{figure}[t]
\setlength{\unitlength}{0.85\textwidth}
\hfill
\begin{picture}(1,0.41)(0,0)
\put(0,0.38){\fixsubfiglabel{a}}
\put(0.42,0.38){\fixsubfiglabel{b1}}
\put(0.75,0.38){\fixsubfiglabel{b2}}
\put(0,-0.35){\includegraphics[width=\unitlength]{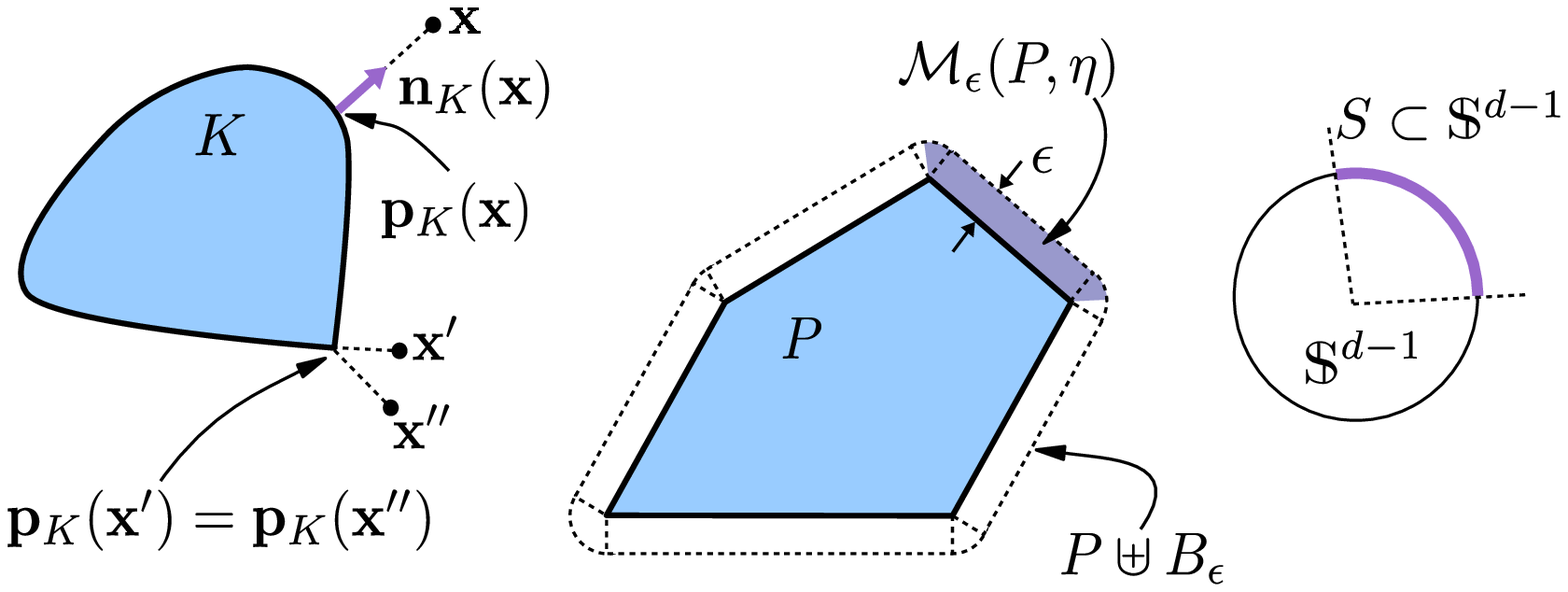}}
\end{picture}
\caption{Construction of a local parallel set. \fixsubfiglabel{a} Definition of the normal field over an arbitrary convex body $K$ at its boundary $\partial K$.
\fixsubfiglabel{b1-b2}
Local parallel set $\mathcal{M}_\epsilon(P,\eta)$ (illustrated for a polytope $P$)
where only points are
considered for which the normal direction is in
a prescribed subset $S$ of the unit sphere. (Here: $\eta=\mathbbm{R}^d \times S$.)}
\label{Fig:Steiner}
\end{figure}

If support measures are
available, then the MT for convex (or more general) bodies are
obtained by integrating tensor functions with respect to these measures. Here we
describe the
approach for convex sets. The idea underlying the introduction of support
measures for convex sets is to generalise
the notion of a parallel set (or ``dilation'', see Fig.~\ref{Fig:Steiner} for $d=2$) of a convex
body $K$ in $d$-dimensional
Euclidean space $\mathbbm{R}^d$ to a suitable
local construction.

A definition of the local parallel set that also applies to convex bodies $K$
without smooth boundary is given in the following: We define
$\mathbf{p}_K(\mathbf{x})$ as the unique point in $K$ which is nearest to a
given point $\mathbf{x}\in\mathbbm{R}^d$. This defines a continuous map
$\mathbf{p}_K(\cdot):\mathbbm{R}^d\to K$, $\mathbf{x}\mapsto
\mathbf{p}_K(\mathbf{x})$. Then
$d_K(\mathbf{x}):=\|\mathbf{x}-\mathbf{p}_K(\mathbf{x})\|$ is the distance from
$\mathbf{x}$ to $K$ and
$\mathbf{n}_K(\mathbf{x}):=(\mathbf{x}-\mathbf{p}_K(\mathbf{x}))/d_K(\mathbf{x}
)$, for $\mathbf{x}\in\mathbbm{R}^d\setminus K$, is
an exterior unit normal of $K$ at the boundary point
$\mathbf{p}_K(\mathbf{x})\in \partial K$ (see Fig.~\ref{Fig:Steiner}).
This construction achieves a definition of nearest points (reminiscent of the
Euclidean distance map \cite{Danielsson:1980}) and surface normals that is also
well-defined for points $\mathbf{x}$ whose nearest point on $K$ is a sharp
corner (where the tangent plane is not well-defined and hence the
conventional differential geometric definition of the surface normal does not
apply).

Now we shall derive a local Steiner formula and support measures, following
ref.~\cite{SchneiderWeil:2008} to obtain local support
measures. The definition of MF and MT then follows directly as a special
case \cite{HugSchneiderSchuster:2008, HugSchneiderSchuster:2008b}. The
intuitive idea underlying the definition of a local parallel set is
described in two steps. First, we specify a region $L\subset \mathbbm{R}^d$ and some
$\epsilon>0$. Then we consider all points $\mathbf{x}\in \mathbbm{R}^d\setminus K$ which have
distance $d_K(\mathbf{x}) $ at most $\epsilon$ from $K$ and for which
$\mathbf{p}_K(\mathbf{x})\in L$. Second, if $K$ has points (such as at sharp
corners or edges) where the exterior surface unit normal vector is not
unique, then it is natural to restrict the points in this local outer
parallel set further by also requiring $\mathbf{n}_K(\mathbf{x})$ to lie in
a prescribed set $S\subset \mathbbm{S}^{d-1}$. As an example, consider the
polytope $P$ in Fig.~\ref{Fig:Steiner}  (b1); for the definition of the
local parallel set (shaded dark) it is necessary to specify which angular
fraction of the wedges 
should be part of the local parallel set. This motivates the definition of
the local parallel set by specification of the spatial region $L$ and by the
subset $S\subset \mathbbm{S}^{d-1}$ of normal directions, conveniently
combined to $\eta = L\times S \subset \mathbbm{R}^d \times
\mathbbm{S}^{d-1}$, where $\mathbbm{S}^{d-1}$ is the $d$-dimensional unit sphere.
This two step procedure can be merged and slightly
extended to the following general definition. 

For given $\epsilon>0$ and $\eta\subset\mathbbm{R}^d\times\mathbbm{S}^{d-1}$,
the local parallel set of $K$ defined by
\begin{equation}
  \label{eq:daniels-def-of-parallel-body}
  \mathcal{M}_\epsilon (K,\eta):=\{\mathbf{x}\in \mathbbm{R}^d\setminus
K:d_K(\mathbf{x})\le
\epsilon,(\mathbf{p}_K(\mathbf{x}),\mathbf{n}_K(\mathbf{x}))\in\eta\}
\end{equation}
contains all points $\mathbf{x}\in \mathbbm{R}^d$ with $0<d_K(\mathbf{x})\le
\epsilon$ such that the pair
$(\mathbf{p}_K(\mathbf{x}),\mathbf{n}_K(\mathbf{x}))\in
\eta$.
The first condition restricts $\mathbf{x}$ to the global outer
parallel set $(K\uplus B_\epsilon)\setminus K$ (see Fig.~\ref{Fig:Steiner} (b1)).
The second condition restricts $\mathcal{M}_\epsilon (K,\eta)$ to those points $\mathbf{x}\in\mathbbm{R}^d$
of the global outer parallel set, where $(\mathbf{p}_K(\mathbf{x}),\mathbf{n}_K(\mathbf{x}))\in \eta$.

 The volume of this local parallel set of a convex body $K$ is
$V_d(\mathcal{M}_\epsilon(K,\eta))$. A fundamental result in integral geometry, known as
the {\em local Steiner formula} \cite{SchneiderBrunnMinkowski,SchneiderWeil:2008},
states that the map $\epsilon\mapsto V_d(\mathcal{M}_\epsilon(K,\eta))$ for all $\epsilon> 0$
($V_d$ is the $d$-dimensional Lebesgue-measure)
is a polynomial of degree $d$,
that is
\begin{equation}\label{eq:locsteiner}
V_d(\mathcal{M}_\epsilon(K,\eta))=\sum_{\nu=0}^{d-1}\epsilon^{d-\nu}\kappa_{d-\nu}\Lambda_\nu(K,\eta),
\end{equation}
where $\kappa_n:=\pi^{n/2}/\Gamma(\frac{n}{2}+1)$  is the volume of an $n$-dimensional unit ball and
 $\Lambda_\nu(K,\eta)$, $\nu=0,\ldots,d-1$, are certain real coefficients that depend on $K$ and $\eta$, but not
 on $\epsilon$.

For eq.~(\ref{eq:locsteiner}) to be true for all $\epsilon>0$, it is
crucial that $K$ is convex \cite{Heveling2004}.
Equation (\ref{eq:locsteiner}) is easily confirmed (and evaluated)
for a convex polytope $P$.
In this case, the set $(K\uplus B_\epsilon)\backslash P$ can be decomposed in an elementary way into wedges over the
faces $F$ of $P$ as indicated by Fig.~\ref{Fig:Steiner} (b1). Let $\mathcal{F}_\nu(P)$ denote the
$\nu$-dimensional faces of the polytope $P$ \footnote{$\mathcal{F}_1$ is the set of edges. In sec.~\ref{sec:algo-for-triangul-bodies}
we use this notion for oriented edges, that is (non-oriented) edges are split into two oriented edges pointing in opposite directions.
Here the set $\mathcal{F}_1$ contains non-oriented edges. Elsewhere it is stated explicitly.} and
\begin{eqnarray} %
&\mathbf{n}(P,F):=\{\mathbf{n}_P(\mathbf{x})\in \mathbbm{S}^{d-1} : \mathbf{p}_P(\mathbf{x}) \in \mathrm{relint}\, F , \mathbf{x}\in\mathbbm{R}^d \backslash P \}
\end{eqnarray} %
 the set of unit normal vectors assigned to $F\in\mathcal{F}_\nu(P)$ and $\mathrm{relint}\,F$ the relative interior of $F$ (i.e.~the interior of $F$ w.r.t. the lowest-dimensional embedding affine hull); see Fig.~\ref{Fig:Illu-npf}.
\def\imagetop#1{\vtop{\null\hbox{#1}}}
\begin{figure}
  \begin{tabular}{llll}
    \fixsubfiglabel{a}  &  \fixsubfiglabel{b} &  \fixsubfiglabel{c}  &  \fixsubfiglabel{d}\\ 
    \imagetop{\includegraphics[width=.31\textwidth]{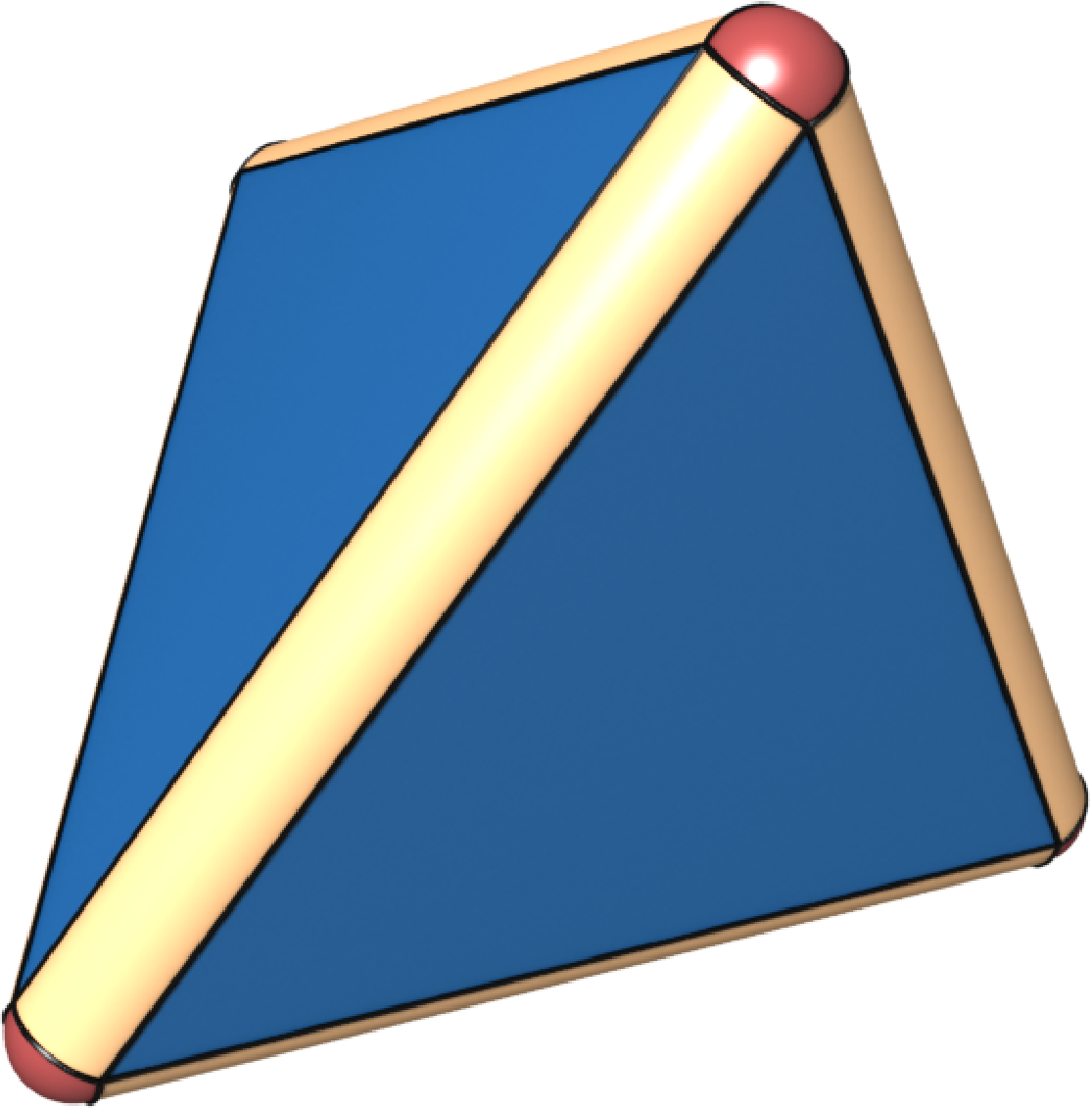}} &
    \imagetop{\includegraphics[width=.2\textwidth]{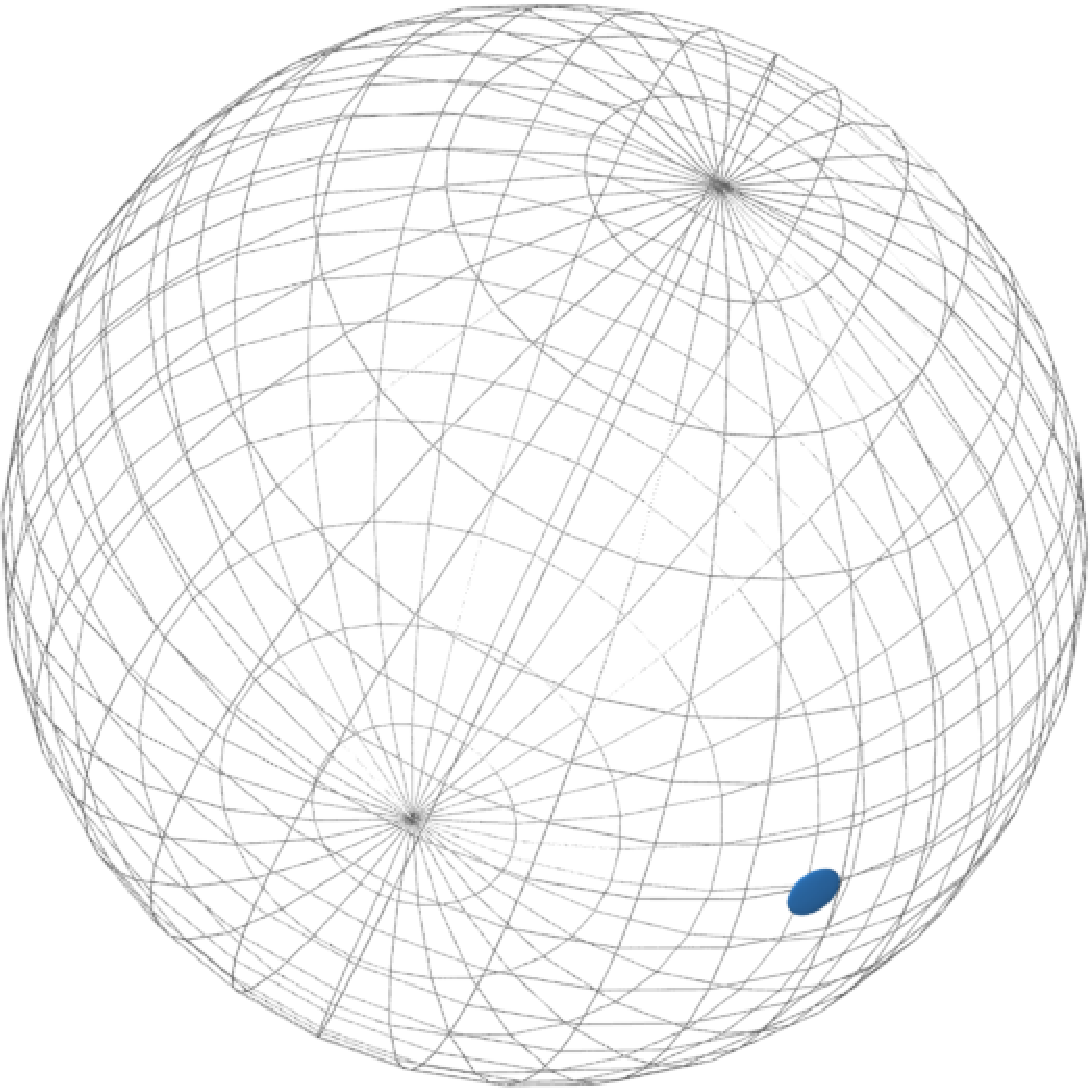}}&
    \imagetop{\includegraphics[width=.2\textwidth]{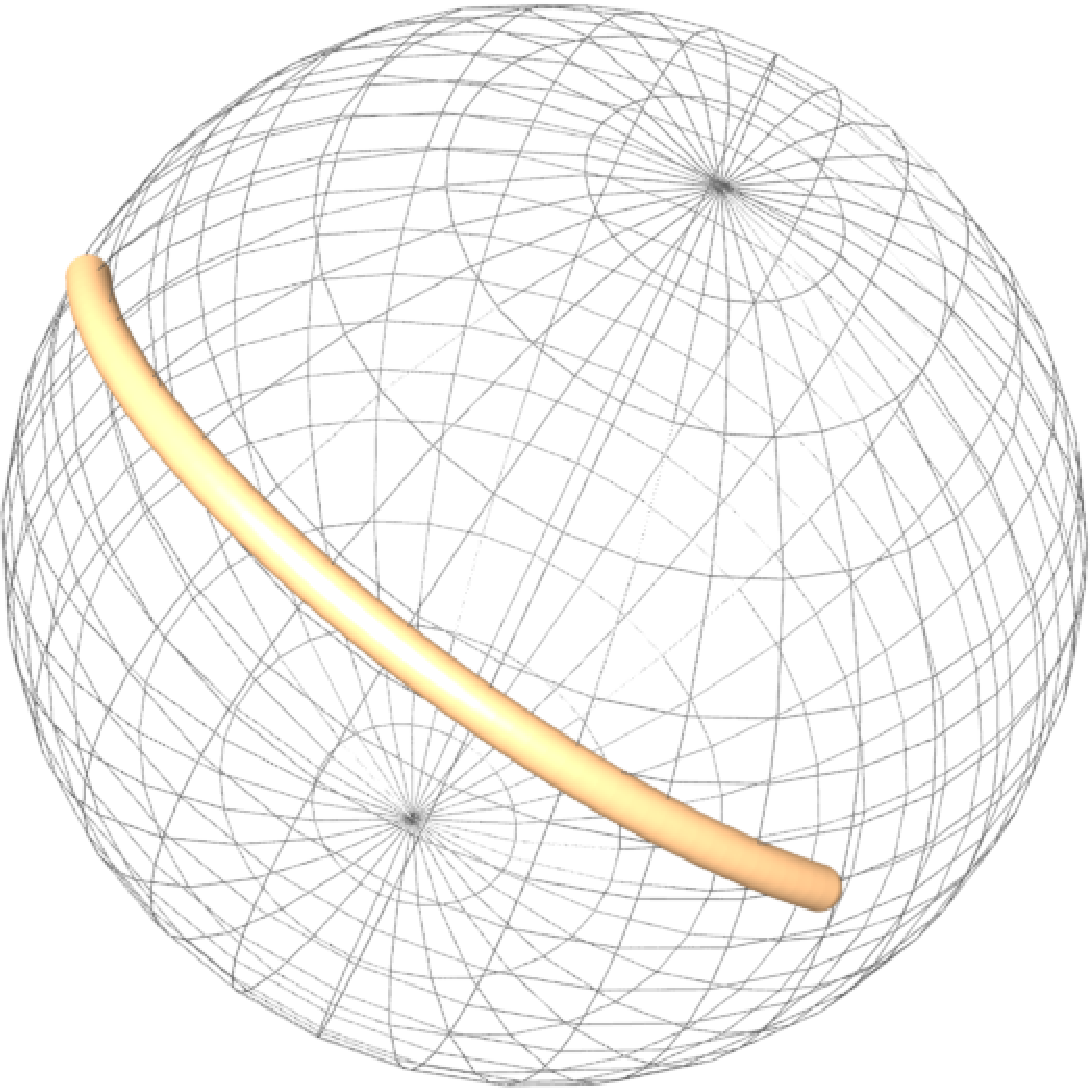}}&
    \imagetop{\includegraphics[width=.2\textwidth]{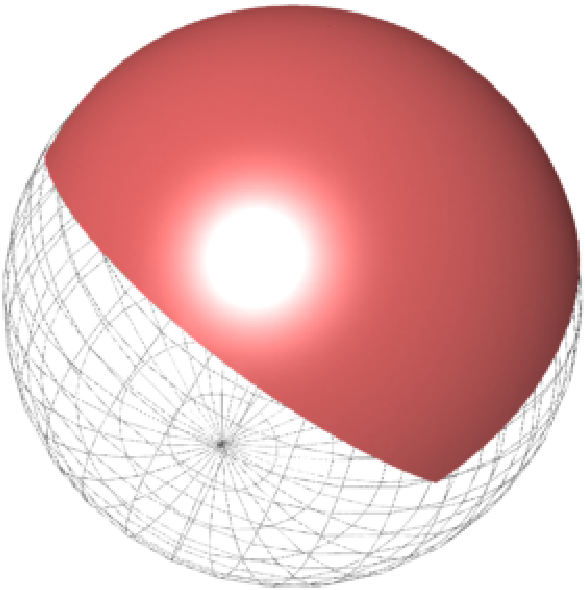}}
  \end{tabular}
\caption{\label{Fig:Illu-npf} \fixsubfiglabel{a} Illustration of the set of normal vectors on the surface of a polytope $P$. \fixsubfiglabel{b}
Normal vector assigned to a facet $F\in\mathcal{F}_2(P)$. \fixsubfiglabel{c} Line segment of normal vectors assigned
to an edge $F\in\mathcal{F}_1(P)$. \fixsubfiglabel{d} Spherical triangle/patch of normal vectors assigned to a vertex $F\in\mathcal{F}_0(P)$.
}
\end{figure}
The contributions to $V_d(\mathcal{M}_\epsilon(K,\eta))$ coming from each of these wedges
\begin{equation}
 \mathcal{M}_\epsilon(P,F):=\mathcal{M}_\epsilon(P,(\mathrm{relint}\,F\times \mathbbm{S}^{d-1}))
\end{equation}
can be calculated by
a simple integration known as Fubini's theorem or Cavalieri's principle
\cite{Cohn:1993,AmannEscher:2009}. This shows that for
$\eta=L\times S$ eq.~(\ref{eq:locsteiner}) holds with
\begin{equation}\label{eq:LocalMeasuresPolytope}
\Lambda_\nu(P,L\times S)=\frac{1}{\omega_{d-\nu}}\sum_{F\in \mathcal{F}_\nu(P)}
\int_{F\cap L}\d \mathcal{H}^\nu\int_{\mathbf{n}(P,F)\cap S}\d \mathcal{H}^{d-\nu-1},
\end{equation}
where $\omega_\nu=\nu\,\kappa_\nu$ is the surface measure of the
$(\nu-1)$-dimensional unit sphere and $\d \mathcal{H}^\nu$ is the Hausdorff measure
of dimension $\nu$ \cite{evans1992measure}.

Since an arbitrary convex body $K$ can be approximated by polytopes,
eq.~(\ref{eq:locsteiner}) can be derived by a continuity argument. In fact,
 $\Lambda_\nu(K,\eta)$ can be expressed as a linear combination of
$V_d(\mathcal{M}_{\epsilon_\mu}(K,\eta))$, for $\epsilon_\mu>0$ pairwise different and $\mu=1,\ldots,d$. One obtains
an invertible $d\times d$-matrix equation (with the matrix entries
$\epsilon_\mu^{d-\nu}\kappa_{d-\nu}$) with $\nu$ running from $0,\ldots,d-1$.
Therefore, the properties of the local parallel volume $V_d(\mathcal{M}_\epsilon(K,\eta))$ (in particular, additivity and weak continuity)  are also available
for $\Lambda_\nu(K,\eta)$ \cite[p. 202]{SchneiderBrunnMinkowski}. In
particular,
$K\mapsto \Lambda_\nu(K,\eta)$ is an additive functional for fixed $\eta$. That is, for convex
bodies $K$ and $K'$
\begin{equation}\label{eq:LocalAdditivity}
 \Lambda_\nu(K \cup K',\eta)=\Lambda_\nu(K ,\eta)+\Lambda_\nu(K',\eta)-\Lambda_\nu(K \cap K',\eta).
\end{equation}
Furthermore, $\eta\mapsto \Lambda_\nu(K,\eta)$
is a non-negative measure for fixed $K$. The latter means that
if $\eta_\mu\subset \mathbbm{R}^d\times\mathbbm{S}^{d-1}$, $\mu\in\mathbbm{N}$,
is a sequence of mutually disjoint (measurable)
sets, then
\begin{equation}\label{Eqn:SigmaAdd}
\Lambda_\nu\left(K,\bigcup_{\mu=1}^\infty\eta_\mu\right)=\sum_{\mu=1}^\infty\Lambda_\nu(K,\eta_\mu).
\end{equation}
This property is called $\sigma$-additivity of $\Lambda_\nu(K,\cdot)$.
Weak continuity of $\Lambda_\nu$ means that for every sequence of convex bodies $K_\mu$ $(\mu\in\mathbbm{N})$, with
$K_\mu\rightarrow K$ and every continuous function $f:\mathbbm{R}^d\times \mathbbm{S}^{d-1}\rightarrow [0,\infty)$
the equation

\begin{equation}
\label{eq_WeakContinuity}
 \lim_{\mu\rightarrow \infty} \int f(\mathbf{x},\mathbf{n}) \,
\Lambda_\nu(K_\mu,\mathrm{d}(\mathbf{x},\mathbf{n}))\nonumber = \int f(\mathbf{x},\mathbf{n}) \,\Lambda_\nu(K,\mathrm{d}(\mathbf{x},\mathbf{n}))
\end{equation}
 holds.
Note that this does not imply $\Lambda_\nu(K_\mu,\eta)\rightarrow \Lambda_\nu(K,\eta)$ as $\mu\rightarrow \infty$
for all measurable sets $\eta\subset \mathbbm{R}^d\times\mathbbm{S}^{d-1}$

In particular, $\Lambda_\nu(K,\cdot)$ can be used to integrate functions
over $\mathbbm{R}^d\times\mathbbm{S}^{d-1}$.
It is plausible that $\Lambda_\nu(K,\cdot)$ is concentrated on the normal
bundle $N(K):=\{(\mathbf{p}_K(\mathbf{x}),\mathbf{n}_K(\mathbf{x}))\in
\partial K\times\mathbbm{S}^{d-1}:\mathbf{x}\in \mathbbm{R}^d\setminus K\}$.  In other
words, $N(K)$ consists of all $(\mathbf{x},\mathbf{n})\in\partial K\times
\mathbbm{S}^{d-1}$ such that $\mathbf{n}$ is an exterior unit normal vector of
$K$ at $\mathbf{x}$. The measures $\Lambda_\nu(K,\cdot)$,
$\nu=0,\ldots,d-1$, are called {\em support measures} and are determined as
coefficient measures of the  Steiner formula, eq.~(\ref{eq:locsteiner}). They are
local versions of the classical MF $W_\nu$, since
$\Lambda_\nu(K,\mathbbm{R}^d\times\mathbbm{S}^{d-1})=V_\nu(K)\propto
W_{d-\nu}(K)$.

If $K$ is sufficiently smooth, then
\begin{equation}\label{smoothrep}
\Lambda_{\nu}(K,\eta)=\frac{\gerdsbinomial{d-1}{\nu}}{\omega_{d-\nu}}
\int_{\partial K}
\mathbf{1}\left\lbrace(\mathbf{x},\mathbf{n}_K(\mathbf{x}))\in\eta \right\rbrace
G_{d-\nu}(\mathbf{x})\, \d A,
\end{equation}
where $G_\nu(\mathbf{x})$ is the
$(\nu-1)$-th (normalised) elementary symmetric function of the principal
curvatures of $\partial K$ at $\mathbf{x}$. That is
in three dimensions, these are $1$, the mean curvature and Gaussian curvature, respectively.
$\mathbf{1}\lbrace\cdot\rbrace$ is the characteristic function, which is
evaluated to one if $\cdot$ is true and $0$ otherwise. For general dimensions and
$\nu\leq d-1$, eq.~(\ref{smoothrep}) holds for all sufficiently smooth convex bodies $K$.

Having introduced the support  measures as local versions of the scalar
MF, it is easy to define the MT
 for a convex body $K$ by
\begin{equation}\label{eq:defMinktens}
\Phi^{r,s}_\nu(K):= \frac{ 1}{r!s!}\frac{\omega_{d-\nu}}{\omega_{d-\nu+s}}
\int_{\mathbbm{R}^d\times\mathbbm{S}^{d-1}}
\mathbf{x}^r\mathbf{n}^s\,\Lambda_\nu\left(K,\d(\mathbf{x},\mathbf{n})\right) ,
\end{equation}
hence we obtain $\Phi^{r,s}_\nu(K)$  by integrating the tensorial function
$\mathbf{x}^r\mathbf{n}^s$ with
respect to the measure $\Lambda_\nu(K,\cdot)$  over
$N(K)\subset\mathbbm{R}^d\times\mathbbm{S}^{d-1}$.
If $K$ is a polytope, for $d=3$ this yields equation (\ref{Tensorforpolytopes}), up to a
different normalisation. If $K$ is smooth, we obtain
eq.~(\ref{Eqn:DefSurfaceTensors}), up to a different normalisation and indexing
scheme, i.e.
\begin{eqnarray}\label{Eqn:MathMT}
&\Phi^{r,s}_\nu(K)=\frac{\gerdsbinomial{d-1}{\nu}}{r!s!\omega_{d-\nu+s}}\int_{\partial K}\mathbf{x}^r\mathbf{n}^s
G_{d-\nu}(\mathbf{x})\, \d A,\nonumber\\
&\Phi^{r,0}_d(K)=\frac{1}{r!}\int_{K} \mathbf{x}^r \mathrm{d}V.
\end{eqnarray}

The notation $\Phi_{\mu,r,s}$ or $\Phi_{\mu}^{r,s}$ for the MT
in eq.~(\ref{eq:defMinktens}) is preferred in some of the mathematical literature and
differs from the notation $W_\nu^{r,s}$ in
eqs.~(\ref{Eqn:DefVolumeTensors}-\ref{Eqn:DefSurfaceTensors}) only by a
different indexing scheme and a different normalisation.  In $\mathbbm{R}^3$,
i.e.~for $d=3$, the functionals $\Phi_\mu^{r,s}$ and $W_\nu^{r,s}$ are related
by
\begin{equation}
W_\nu^{r,s}(K)=C\,\Phi_{d-\nu}^{r,s}(K),\quad \mathrm{  with}\quad  C:=\frac{r!s!\omega_{\nu+s}}{d\gerdsbinomial{d-1}{\nu-1}},
\end{equation}
for $\nu=1,\ldots,d$, and
\begin{equation}
W_0^{r,0}(K)=r!\,\Phi_{d}^{r,0}(K).
\end{equation}

The additivity and continuity properties of the support measures immediately
yield the corresponding properties of the MT. This approach also shows that if it is
possible to define support measures for a class of sets, then
the corresponding tensor valuations can be defined by eq.~(\ref{eq:defMinktens}).

Since the theory  of support measures is well-developed
\cite{SchneiderWeil:2008,SchneiderBrunnMinkowski}, the
measure theoretic approach outlined above has some advantages over the differential geometric approach.

As a simple illustration, let us explain why $W_{d-\nu}^{1,1}$ is translation
invariant for $\nu=0,\ldots,d-1$. Observe that by translation covariance of the
support measures
\begin{eqnarray}
&\int_{\mathbbm{R}^d\times\mathbbm{S}^{d-1}}
\mathbf{x}\mathbf{n}\,\Lambda_\nu\left(K\uplus\mathbf{t},\d(\mathbf{x},\mathbf{n
})\vphantom{\int}\right)\nonumber\\
&=\int_{\mathbbm{R}^d\times\mathbbm{S}^{d-1}}
(\mathbf{x}+\mathbf{t})\mathbf{n}\,\Lambda_\nu\left(K,\d(\mathbf{x},\mathbf{n}
)\vphantom{\int}\right)\nonumber\\
&=\int_{\mathbbm{R}^d\times\mathbbm{S}^{d-1}} \mathbf{x}\mathbf{n}\,
\Lambda_\nu\left(K,\d(\mathbf{x},\mathbf{n})\vphantom{\int}\right)+\mathbf{t}\int_{\mathbbm{R}^d\times\mathbbm{S}^{d-1}}
\mathbf{n}\,\Lambda_\nu\left(K,\d(\mathbf{x},\mathbf{n})\vphantom{\int}\right).\label{
eq:trans-cov-supp-meas}
\end{eqnarray}

It is a basic property of the measures $\Lambda_\nu(K,\cdot)$ that they are
centred at the origin
in the sense that $\int \mathbf{n}\,\Lambda_\nu(K,\d(\mathbf{x},\mathbf{n}))=0$ \cite{Mueller:1953}, which
yields the assertion.

A  natural and useful extension
that is suggested by general measure theory is to introduce local tensor valuations by restricting
the integration on the right-hand side of eq.~(\ref{eq:defMinktens}) to measurable subsets of
$\mathbbm{R}^d\times\mathbbm{S}^{d-1}$.

\section{Bodies Bounded by Triangulated Surfaces}
\label{sec:algo-for-triangul-bodies}

We describe an exact algorithm for all independent scalar, vectorial and rank-2
MT of bodies bounded by piece-wise linear
(i.e.~triangulated) surfaces. Henceforth such bodies, convex or non-convex,
are called {\em polytopes} $P$ and their bounding surface is the {\em triangulation} 
$\mathcal{F}$. Triangulations are commonly used as discrete approximations of smooth surfaces. The continuity of the MF and MT guarantee
the convergence of the formula for the triangulations to the MF and MT of the smooth body.

The formulae are derived for convex bodies with
triangulated bounding surfaces by considering parallel bodies $P_\epsilon$ of
convex polytopes $P$ (that is $P_\epsilon$ has a continuous normal fields and
finite curvatures for $\epsilon> 0$ and a well defined limit as $\epsilon
\rightarrow 0$). By application of the
additivity relation these formulae are then shown to be valid also for bodies
that are not convex. As the key results of this article---explicit formulae for
the computation of MT of convex and non-convex polytopes---are
summarised in table \ref{Tab:MinkowskiValuationsForTriangulations}.

\begin{figure*}[t!]
\includegraphics[width=\textwidth]{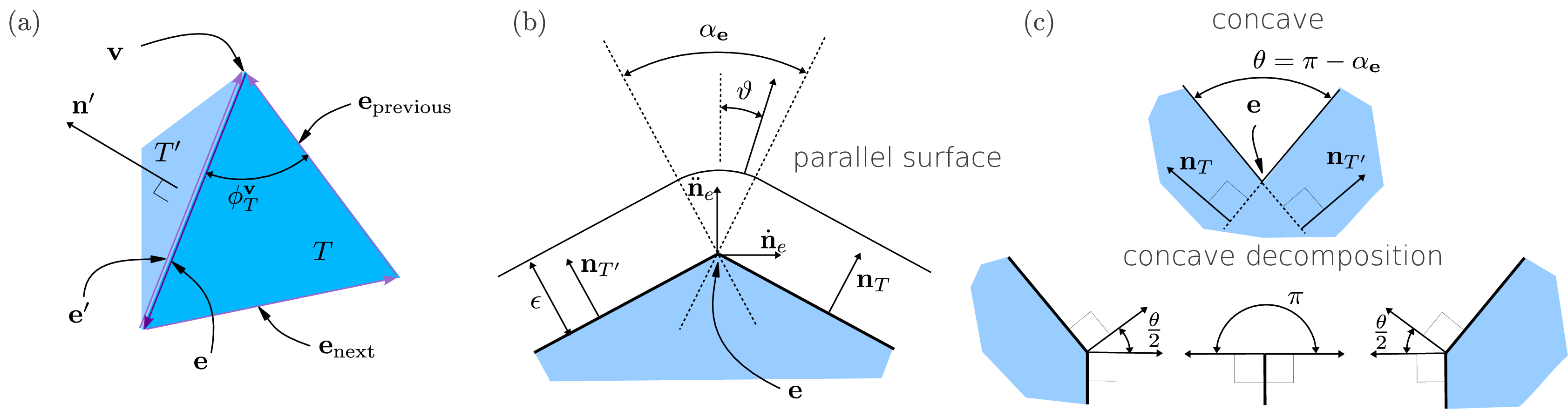}
\caption{
\fixsubfiglabel{a} Definition of geometric properties of a triangulated
surface $\partial P$ with doubly connected edge list (DCEL).
Each edge $\mathbf{e}$ is oriented and uniquely assigned to a triangle $T$.
The counter oriented edge to $\mathbf{e}$ is denoted $\mathbf{e}'$ and 
assigned
to the adjacent triangle $T'$. An oriented edge
 $\mathbf{e}$ is unambiguously assigned to the previous edge
$\mathbf{e}_\mathrm{previous}$
and the next edge $\mathbf{e}_\mathrm{next}$. The
 normal vector $\mathbf{n}_T$ is defined to point out of the body $K$, i.e.~
$\mathbf{n}_T=(\mathbf{e}_\mathrm{previous}\times
 \mathbf{e})/| \mathbf{e}_\mathrm{previous}\times \mathbf{e}|$. The
 angle between two edges of
 triangle $T$ at vertex $\mathbf{v}$ is denoted $\phi_T^{\mathbf{v}}$.  \fixsubfiglabel{b}
Cross-sectional view along an oriented edge $\mathbf{e}$. The normal vectors
$\mathbf{n}_T$ and $\mathbf{n}_{T'}$ of the
 triangle $T$ (that contains $\mathbf{e}$) and $T'$ span the angle
$\alpha_{\mathbf{e}}\in(-\pi,\pi]$. A concave edge has a negative angle
 $\alpha_{\mathbf{e}}$. The figure also shows the definition of the local
coordinate system used for the computation of
 $W_2^{0,2}$. The basis vectors $\dot{\mathbf{n}}_\mathbf{e},
\ddot{\mathbf{n}}_{\mathbf{e}}$ and $\hat{\mathbf{e}}$ are
 defined as $\hat{\mathbf{e}}=\mathbf{e}/\vert \mathbf{e} \vert$,
$\ddot{\mathbf{n}}_\mathbf{e} =
 (\mathbf{n}_{T}+\mathbf{n}_{T'})/\vert\mathbf{n}_{T}
+\mathbf{n}_{T'}
\vert$ and $\dot{\mathbf{n}}_\mathbf{e}=\ddot{\mathbf{n}}_\mathbf{e}\times\hat{\mathbf{e}}$.
\fixsubfiglabel{c} Subdivision of a body $P$ along a concave edge $\mathbf{e}$ to yield locally convex bodies.
}
\label{Fig:NotationTriangulation}
\end{figure*}

Consider a polytope $P$ in $\mathbbm{R}^3$ with piece-wise linear bounding surface
$\partial P \equiv\mathcal{F}$. Without loss of generality the linear facets may
be assumed to be
triangles.\footnote{It is an important consequence of the additivity relation
that the MT (in contrast to e.g.~the texture tensor) do not
change if flat polygonal facets are broken up into triangles. This is evidently
also true for the algorithmic implementation described here.} The set of all
triangular patches of $\partial P$ is $\mathcal{F}_2$, the set of oriented edges
is $\mathcal{F}_1$ and the set of vertices is $\mathcal{F}_0$. We assume a
\emph{doubly
connected edge list} (DCEL \cite{berg:2000}, also called \emph{half edge data
structure} \cite{cgal}), that is,
every edge which is shared between two triangles $T$ and $T'$ is a double-edge
consisting of two oriented edges $\mathbf{e}$ (being part of $T$) and
$\mathbf{e}'$ (part of $T'$), constituting an unambiguous assignment of each
edge to a triangle. Each oriented edge is assigned to its previous edge
$\mathbf{e}_\mathrm{previous}$ and its next edge
$\mathbf{e}_\mathrm{next}$. The remaining ambiguity in the edge orientation is
lifted by requiring the triangle normals
$\mathbf{n}_T=(\mathbf{e}_\mathrm{previous}\times \mathbf{e})/|
\mathbf{e}_\mathrm{previous}\times \mathbf{e}|$ to point out of the body
$P$.
Thus we can uniquely assign to each oriented edge $\mathbf{e}$ a triangle $T$
with vertices $\mathbf{v}_1$, $\mathbf{v}_2$ 
$\mathbf{v}_3$ (see Fig.~\ref{Fig:NotationTriangulation}) and its normal vector
$\mathbf{n}_T$.

The parallel body construction is illustrated by Fig.~\ref{Fig:Body-illu}~(b).
For an arbitrary body $P$ the parallel body $P_\epsilon$ with thickness
$\epsilon > 0$ is defined as 
$P_\epsilon:=P\uplus B_\epsilon$.
 For a convex polytope $P$ (whose bounding surface has a discontinuous normal field) the
bounding surface $\partial P_\epsilon$ has a
continuous normal field. The curvatures are patch-wise constant: $G_2=G_3=0$ on
the planar patches, $G_2=(2\epsilon)^{-1}$ and $G_3=0$ on the
cylindrical patches corresponding to polygon edges, and $G_2=1/\epsilon$ and
$G_3=1/\epsilon^2$ on the spherical patches corresponding to polytope vertices.
For convex polytopes, the MT are defined as the surface
integrals of eq.~(\ref{Eqn:DefSurfaceTensors}) evaluated on $\partial
P_\epsilon$ in the limit $\epsilon \rightarrow 0$. The result thus obtained is 
equivalent with eq.~(\ref{eq:defMinktens}), see also eq.~(\ref{Tensorforpolytopes}).

\subsection{\label{subsec:volume}Volume $W_0$ }

\begin{table*}[t]
\centering
\begin{tabular*}{\textwidth}{@{\extracolsep{\fill}}  lll }
\hline
\multicolumn{3}{c}{scalar measures}\\[0.1cm]
   $W_0$ &
    $\int_K \d V$  & $\frac 13 \Sum{T\in\mathcal{F}_2}{}\langle \mathbf{C}_T,
\mathbf{n}_T \rangle \vert T\vert$
    \\[0.40cm]
    $W_1$ &
    $\frac{1}{3}\int_{\partial K} \d A$ & $\frac{1}{3}\Sum{T\in\mathcal{F}_2}{}
\vert T\vert$
    \\[0.40cm]
    $W_2$ &
    $\frac{1}{3}\int_{\partial K} G_2 \d A$ &
$\frac{1}{12}\Sum{\mathbf{e}\in\mathcal{F}_1}{} \vert\mathbf{e}\vert
\,\alpha_{\mathbf{e}}$
    \\[0.40cm]
    $W_3$ &
    $\frac{1}{3}\int_{\partial K} G_3 \d A$
&$\frac{1}{3}\Sum{\mathbf{v}\in\mathcal{F}_0}{}(2\pi-\Sum{T\in\mathcal{F}
_2(\mathbf{v})}{}\phi_{T}^{\mathbf{v}})$
    \\*[0.40cm]
\hline
\multicolumn{3}{c}{vectorial measures}\\[0.1cm]
    $(W_0^{1,0})_i$
    &
    $\int_K \mathbf{x}_i \d V$
    &
    $\Sum{T\in\mathcal{F}_2}{} \,(I_{T})_{ik} (n_{T})_{k}$, see
sec. \ref{sec:w010deriv}
    \\[0.40cm]
    $(W_1^{1,0})_i $&$\frac{1}{3}\int_{\partial K} \mathbf{x}_i \d A$ & $
    \frac{1}{3}\Sum{T\in\mathcal{F}_2}{}\vert T\vert (\mathbf{C}_{T})_{i}$ 
    \\[0.40cm]
    $(W_2^{1,0})_i $&$\frac{1}{3} \int_{\partial K} G_2\,\mathbf{x}_i \d A$ & 
    $\frac{1}{12}\Sum{\mathbf{e}\in\mathcal{F}_1}{} \vert \mathbf{e}\vert
\alpha_{\mathbf{e}} (\mathbf{C}_{\mathbf{e}})_{i}$
    \\[0.40cm]
    $(W_3^{1,0})_i $&$\frac{1}{3} \int_{\partial K} G_3\,\mathbf{x}_i \d A$ & $

\frac{1}{3}\Sum{\mathbf{v}\in\mathcal{F}_0}{}(2\pi-\Sum{T\in\mathcal{F}
_2(\mathbf{v})}{}\phi_{T}^{\mathbf{v}})\mathbf{v}_i$ 
    \\[0.40cm]
    \hline
    \multicolumn{3}{c}{tensorial measures (rank two)}\\[0.1cm]
    $(W_0^{2,0})_{ij}$ & $\int_K \mathbf{x}_i\mathbf{x}_j \d V $
    &
    $\Sum{T\in\mathcal{F}_2}{} \,(J_{T})_{ijk} (n_{T})_{k}$, see
sec. \ref{sec:w020deriv}
    \\[0.40cm]
    $(W_1^{2,0})_{ij}$
    &
    $\frac{1}{3}\int_{\partial K} \mathbf{x}_i\mathbf{x}_j \d A $
    &
    $\frac{1}{3}
        \Sum{T\in\mathcal{F}_2}{}
         (I_T)_{ij}  $ 
    \\[0.40cm]
    $(W_2^{2,0})_{ij}$
    &
    $\frac{1}{3}\int_{\partial K} G_2\,\mathbf{x}_{i}\mathbf{x}_{j} \d A $ 
    &
    $\frac{1}{36}\Sum{\mathbf{e}\in\mathcal{F}_1}{} \alpha_{\mathbf{e}} |\mathbf
e| \cdot \left(

(\mathbf{v}_{1}^2)_{ij}+(\mathbf{v}_{1}\mathbf{v}_{2})_{ij}+(\mathbf{v}_{2}^2)_{
ij}
        \right) $
    \\[0.40cm]
    $(W_3^{2,0})_{ij}$ & $\frac{1}{3}\int_{\partial K}
G_3\,\mathbf{x}_i\mathbf{x}_j \d A $ &
$\frac{1}{3}\Sum{\mathbf{v}\in\mathcal{F}_0}{}\left(2\pi-
    \Sum{T\in\mathcal{F}_2(\mathbf{v})}{}\phi_{T}^{\mathbf{v}}
    \right)(\mathbf{v}^2)_{ij}$
    \\
    $(W_1^{0,2})_{ij}$ & $\frac {1}{3}\int_{\partial K}
\mathbf{n}_i\mathbf{n}_j \d A$&$
    \frac{1}{3}\Sum{T\in\mathcal{F}_2}{}\vert T\vert\,(\mathbf{n}_{T}^2)_{ij}$
    \\[0.40cm]
    $(W_2^{0,2})_{ij}$ & $\frac 13\int_{\partial K} G_2\,
\mathbf{n}_i\mathbf{n}_j \d A$&$
    \frac{1}{24}\Sum{\mathbf{e}\in\mathcal{F}_1}{}\vert\mathbf{e}\vert
    \left( (\alpha_\mathbf{e}+\sin\alpha_\mathbf{e})({\ddot{\mathbf{n}}_\mathbf{e}}^2)_{ij}+
    (\alpha_\mathbf{e}-\sin\alpha_\mathbf{e})(\dot{\mathbf{n}}_\mathbf{e}^2)_{ij} \right)$
    \\[0.40cm]
    \hline
\end{tabular*}
\begin{minipage}{\textwidth}
\begin{spacing}{0.8}
{
\vspace*{0.2cm}
\footnotesize
\emph{Second column:} MF and MT for bodies with smooth boundary  $\partial K$. The mean and Gaussian
curvatures on $\partial K$ are $G_2$ and $G_3$, respectively. \emph{Third column:}
 MF and MT for a triangulation.
The
set of facets of the triangulation $\mathcal{F}$ is
$\mathcal{F}_2$, the set of
oriented edges is $\mathcal{F}_1$ (in DCEL structure, see text) and the set
of
vertices $\mathcal{F}_0$. The subset of triangles that contain the vertex
$\mathbf{v}$ is denoted by $\mathcal{F}_2(\mathbf{v})$. The nomenclature for
triangulated surfaces is defined in Fig. \ref{Fig:NotationTriangulation}. The vertices of an edge $\mathbf{e}$ or a triangle $T$ are denoted
$\mathbf{v}_1,\mathbf{v}_2$ and $\mathbf{v}_3$, respectively. $\vert T\vert$ is
the area of
$T\in\mathcal{F}_2$, $\mathbf{C}_T$ its centre point $
(\mathbf{v}_1+\mathbf{v}_2+\mathbf{v}_3)/3$ and the tensors $I_T$
and $J_T$ are given in eqs.~(\ref{eq:ITtensor}) and (\ref{eq:JTtensor}) and table \ref{tab:w0utilityfunctions}.
 $\mathbf{C}_\mathbf{e}$ is the centre point of edge $\mathbf{e}$,
$\mathbf{C}_\mathbf{e}=(\mathbf{v}_1+\mathbf{v}_2)/2$. $i,j,k\in\lbrace
x,y,z\rbrace$, 
and $\vert e\vert$ its length. The symbol $\langle\cdot,\cdot\rangle$ denotes the scalar product. $\alpha_e$ is the dihedral angle across edge $\mathbf{e}$, see sec.~\ref{subsec:surf-area-int-mean-curv}. $\mathbf{n}_T$ is the normal of triangle $T$, pointing out of the body, see Fig.~\ref{Fig:NotationTriangulation}. The jump angles $\phi_T^{\mathbf{v}}$ are defined in sec.~\ref{subsec:int-gauss} and Fig.~\ref{Fig:GaussianCurvatureAtAVertex}, and the quantities $\ddot{\mathbf{n}}_\mathbf{e}$ and $\dot{\mathbf{n}}_\mathbf{e}$ below eq.~(\ref{eq:w202}).
}
\end{spacing}
\end{minipage}
\caption{MF and MT in 3D of body $K$ with smooth boundary $\partial K$ and a body $P$ bounded by a
triangulated surface $\partial P$. }
\label{Tab:MinkowskiValuationsForTriangulations}
\end{table*}

The calculation of the volume of a polytope $P$ can be transformed into a
surface integral by Gauss' law, see eq.~(\ref{eq:GaussTheoremFromLinDep})
\cite{Eberly:2004}. With
$\mathrm{div}\, \mathbf{x}=\mathrm{div}
(x,y,z)^t=3$ one obtains
\begin{eqnarray}
W_0(K)&=\int_P \d V = \frac{1}{3}\int_P \mathrm{div}\, \mathbf{x} \d V =
\frac{1}{3}\int_{\partial P} \langle \mathbf{x}, \d\mathbf{A}\rangle \nonumber\\
&=\mathrm{tr}\, \frac{1}{3}\int_{\partial P}  \mathbf{x}\mathbf{n}\,
\mathrm{d}A= \mathrm{tr}\, W_{1}^{1,1}
\label{eq:vol-by-surf-integral-ana}
\end{eqnarray}
where $\d\mathbf{A}=\mathbf{n}\d A$ denotes the oriented infinitesimal area
element and $\langle\cdot,\cdot\rangle$ the scalar product.

\subsection{\label{subsec:surf-area-int-mean-curv}Surface area $W_1$ and integral mean curvature $W_2$}

The surface integral is a sum over triangles and is easily evaluated yielding
the formulae in Tab.~\ref{Tab:MinkowskiValuationsForTriangulations}. This result
is independent whether $P$ is convex or not.
The surface area $W_1(P)$ of $\partial P$ is simply the sum of triangle areas.

\begin{figure}[t]
\hfill \includegraphics[width=0.75\textwidth]{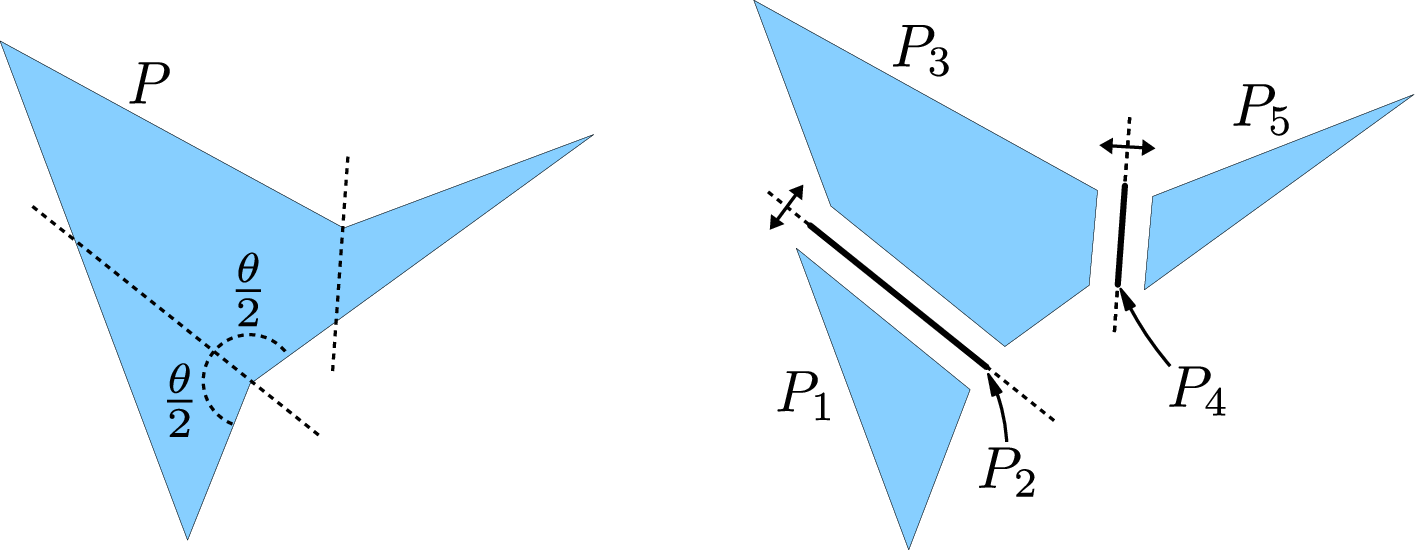}
\caption{\label{fig:subdivision-nonconvex-body-2d} Subdivision of a non-convex body into convex sub-bodies, $P=P_1\cup P_2
\cup P_3 \cup P_4 \cup P_5$. Note that for the computation of MT
the segments $P_2$ and $P_4$ need to be taken into account, though their
volume measure is 0: $W_0(P_2)=W_0(P_4)=0$.}
\end{figure}

Expressing $W_2(P)$ as the limit of vanishing parallel distance $\epsilon$ of
$W_2(P_\epsilon)$ of the parallel body $P_\epsilon$,
$W_2(P)=\lim_{\epsilon\rightarrow 0} W_2(P_\epsilon)$, the contributions of
facets vanish because the mean curvature of a flat face is zero. The
contribution of the spherical patches $S$ corresponding to vertices vanishes because
the integral over a spherical patch $S$ can be parametrised in spherical coordinates
by $\int\int_S \frac{1}{\epsilon} \, \epsilon^2\sin\theta\,d\varphi\,\d\theta $ which vanish as
$\epsilon \rightarrow 0$. The remaining contribution of the edges is located at
the cylindrical patches of $\partial P_\epsilon$ and is given in polar
coordinates by

\begin{equation}
\label{Eqn:W2OfAnEdge}
W_2(P) = \frac{1}{3}\lim_{\epsilon\rightarrow 0} \sum_{{\bf e}\in\mathcal{F}_1} \frac{1}{2}
\int_0^{|{\bf e}|} \d l \int_{-\alpha_{\bf e}/2}^{\alpha_{\bf e}/2} \frac{1}{2\epsilon} \epsilon
\d\vartheta  = 
\sum_{{\bf e}\in\mathcal{F}_1} |{\bf e}|\,\frac{\alpha_{\bf e}}{12} ,
\end{equation}

where $|{\mathbf e}|$ is the length of edge ${\mathbf e}$ and $\alpha_{\mathbf
e}$ the dihedral angle, i.e.~the angle between the surface normals of
the two facets adjacent to ${\mathbf e}$. For a convex body, all edges have a dihedral angle
$0\leq \alpha_\mathbf{e}\leq \pi$; see also
 Fig.~\ref{Fig:NotationTriangulation}. Note that $\mathcal{F}_1$ is the set of
oriented edges, i.e.~the edge shared by
 two triangles is represented by two distinct oriented edges, which explains the factor $1/2$ in front of the integral.\\

Eq.~(\ref{Eqn:W2OfAnEdge}) remains valid even if $P$ is not convex, as is shown
by exploiting additivity: A non-convex polytope $P$ can be decomposed into a set
of convex polytopes by cutting along the {\em symmetric bisector planes} of
all concave edges (that is, $-\pi< \alpha_\mathbf{e}< 0$), see Fig.~\ref{fig:subdivision-nonconvex-body-2d}. For a
concave edge $\mathbf{e}$, the symmetric bisector plane is the plane that is
spanned by $\mathbf{e}$ and the average of the facet normals of the two facets
adjacent to $\mathbf{e}$. By adding the contributions of all resulting convex
bodies using the additivity relationship eq.~(\ref{eq:additivity}), as outlined
in Fig.~\ref{Fig:NotationTriangulation} (c), one obtains the validity of
eq.~(\ref{Eqn:W2OfAnEdge}) for non-convex triangulated bodies. The sign of the
dihedral angle $\alpha_\mathbf{e}\in (-\pi,\pi]$ determines if the edge is
convex ($\alpha_\mathbf{e}>0$) or concave ($\alpha_\mathbf{e}<0$).\\

\subsection{\label{subsec:int-gauss}Integral Gaussian curvature $W_3$ (Euler index $\chi$)}

As the point-wise Gaussian curvature $G_3$ on cylinders and flat facets
vanishes, only vertices of the triangulation (and their corresponding spherical
patches on the parallel body) contribute to $W_3$. For both a convex and
a non-convex polytope  $P$ the point-wise Gaussian curvature $G_3$ and the
integrated Gaussian curvature $W_3$ can be calculated by the well-known simple
sum over angle deficits at surface vertices in
eq.~(\ref{eq:local-Gauss-curvature-polyhedra}), derived below, and also given in
\cite{KrsekLukacsMartin:1998,MeyerDesbrunSchroederBarr:2003}. The non-convex
case is treated by exploiting additivity.\\

\begin{figure}[t]\centering
\includegraphics[width=\textwidth]{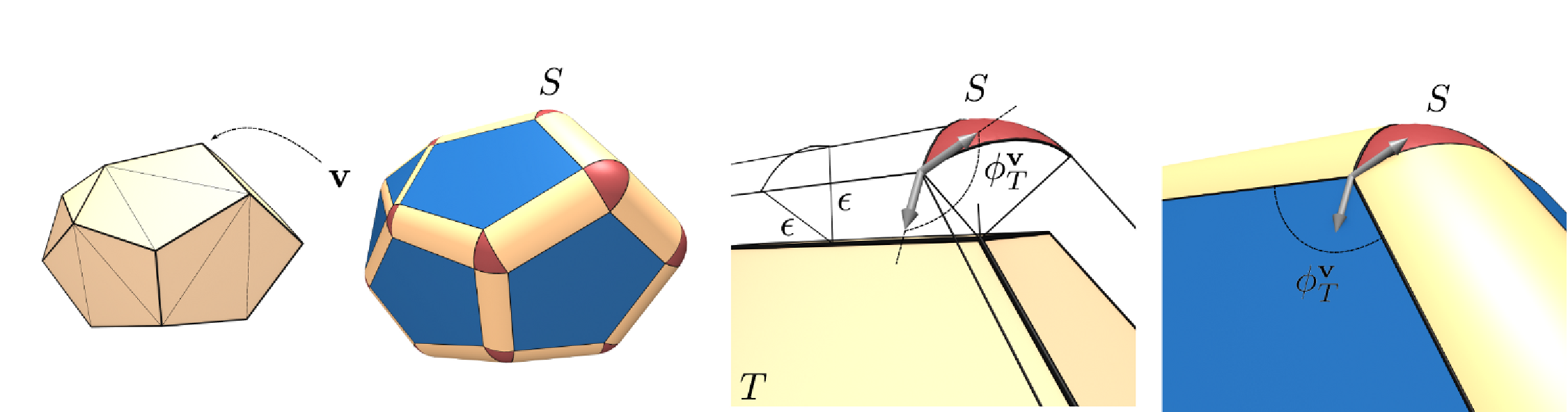}
\caption{Sketch of a vertex $\mathbf{v}$ with a spherical patch $S$
 of the parallel surface $\partial P_\epsilon$. The interior angle in the
triangle $T$ adjacent to $\mathbf{v}$ is denoted $\phi_T^{\mathbf{v}}$.
$S$ is a spherical polygon. The jump angles coincide with the interior angles of the triangles.
}
\label{Fig:GaussianCurvatureAtAVertex}
\end{figure}

\begin{figure}\centering
\includegraphics[width=.43\textwidth]{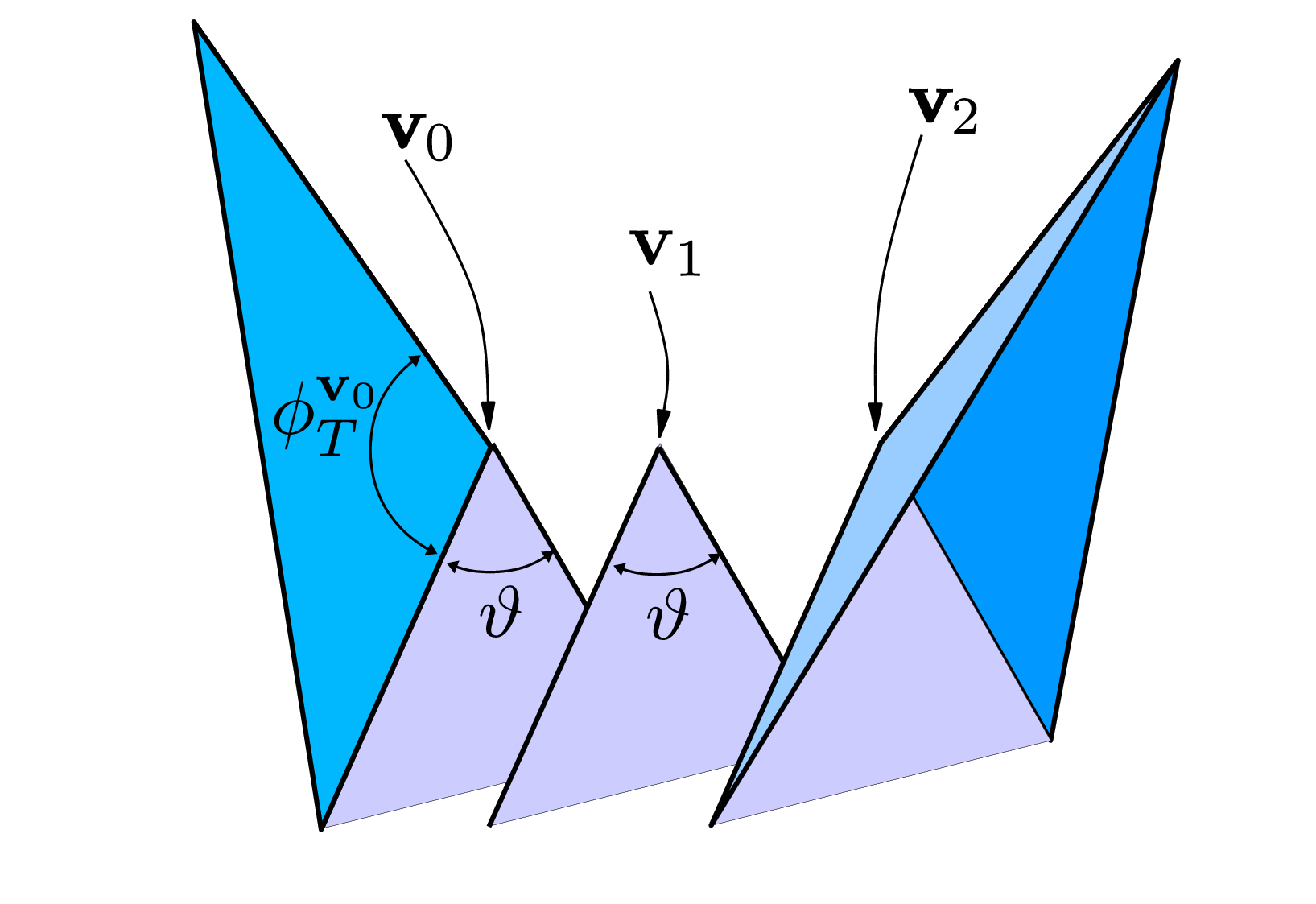}
\caption{The Gaussian curvature $G_3$ at a saddle vertex is obtained by a
virtual decomposition of $P$ at $\mathbf{v}$ into
three polytopes with vertices $\mathbf{v}_\mu$, $\mu=1,2,3$ using
the additivity property of MT.}
\label{FigGaussianCurvatureAtAVertexDecompo}
\end{figure}

The Gaussian curvature contribution of the vertices $\mathbf{v}\in\mathcal{F}_0$
is derived by the Gauss-Bonnet-formula
\begin{equation}
\int_S G_3 \d A = 2\pi-\sum_{T\in\mathcal{F}_2(\mathbf{v})} \phi_T^{\mathbf{v}}-\int_{\partial S} k_g \d s,
\label{Eqn:GaussBonnetFormula}
\end{equation}
where $\mathcal{F}_2(\mathbf{v})$ is the subset of triangles adjacent to vertex
$\mathbf{v}$ and $S$ denotes the spherical patch on the parallel surface $\partial
P_\epsilon$. For all $\epsilon >0$ each spherical cap $S\subset \partial P_\epsilon$
can be uniquely assigned to one vertex $\mathbf{v}$; $\partial
S$ its oriented boundary curve and $k_g$ the geodesic curvature along $\partial
S$. At the corners of $\partial S$, the discontinuity of the tangent vectors is
characterised by {\em jump angles} $\phi_T^{\mathbf{v}}$ (see
Fig.~\ref{Fig:GaussianCurvatureAtAVertex}) which for all $\epsilon >0$ coincide with the
interior angles of the triangle $T$ at $\mathbf{v}$ \cite{Chavel:1996},
see Fig.~\ref{Fig:GaussianCurvatureAtAVertex}. The geodesic curvature $k_g$ vanishes almost everywhere along
$\partial S$, because $\partial S$ are great circle arcs on the spherical patch
and the adjacent cylindrical patch and are thus geodesics; hence the integral $\int_{\partial S}
k_g \d s$ vanishes.

As a consequence, $\int_{S} G_3 \d  A$ is constant for all $\epsilon>0$.
Equation (\ref{Eqn:GaussBonnetFormula}) therefore yields a definition and an
explicit formula for $W_3(P)$ as a sum of the local contribution $w_3(P,\mathbf{v})$
 at a vertex $\mathbf{v}$

\begin{equation}
W_3(P)=\frac{1}{3}\sum_{\mathbf{v}\in\mathcal{F}_0} w_3(P,\mathbf{v})=\frac{1}{3}\sum_{\mathbf{v}\in\mathcal{F}_0} \left(
2\pi-\Sum{T\in\mathcal{F}_2(\mathbf{v})}{}\phi_{T}^{\mathbf{v}}\right).
\label{eq:local-Gauss-curvature-polyhedra}
\end{equation}

At a concave vertex $\mathbf{v}$, a polytope $P$ can always be decomposed into
three separate bodies (one of vanishing volume) that have convex vertices in lieu of $\mathbf{v}$. It is
easy to validate eq.~(\ref{eq:local-Gauss-curvature-polyhedra}) at concave vertices by using the
additivity relation in eq.~(\ref{eq:additivity}), see Fig.~\ref{FigGaussianCurvatureAtAVertexDecompo}.

\subsection{Centre of mass $W_0^{1,0}/W_0$ and curvature centroids $W_\nu^{1,0}/W_\nu$\label{sec:w010deriv}}

The Minkowski vector $W_0^{1,0}$ corresponds to the centre of mass of $P$
multiplied with its volume (assuming $P$ is homogeneously filled with material of
constant density.) The components of this
vector may be computed by transforming the volume integral into a surface
integral using Gauss' theorem
\begin{eqnarray}
    \left(W_0^{1,0}(P)\right)_{i} = \Int{P}{} \mathbf{x}_i \d V= \Int{\partial
P}{} \langle \mathbf{f}_i ,  \d  \mathbf{A} \rangle
\end{eqnarray}
with functions $\mathbf{f}_i$ that satisfy $\mathrm{div}\mathbf{f}_i = \mathbf{x}_i$.
The vector-valued auxiliary function $\mathbf{f}_i$ can be chosen for each index
$i$ independently and the index $i$ denotes the index of $W_0^{1,0}$, which is
evaluated.
For the particular choice of $\mathbf{f}_i$ given in table
\ref{tab:w0utilityfunctions}, this can be explicitly written as
\begin{equation}
    (W_0^{1,0}(K))_{i} = \Sum{T\in\mathcal{F}_2}{}
                           \Int{T}{} \mathbf{x}_i \mathbf{x}_k
(\mathbf{n}_{T})_{k} \d  A   \nonumber
                        = \Sum{T\in\mathcal{F}_2}{}
                             (I_{T})_{ik} (\mathbf{n}_{T})_{k}
    \label{eq:skw010intg}
\end{equation}
with $k$ as listed in table \ref{tab:w0utilityfunctions}.
($k$ is not a summation index.) $\vert T\vert$ is the surface area of $T$.
The $I_T$ in eq.~(\ref{eq:skw010intg}) are tensorial integrals over the
individually
parametrised triangles with the three vertices $\mathbf{v}_\mu$, $\mu=1,2,3$
\begin{eqnarray}\label{eq:ITtensor}
    I_{T} &= 2\vert T\vert  \Int 01 \d a \Int 0{1-a} \d b
\left[\mathbf{v}_{1}+a(\mathbf{v}_{2}-\mathbf{v}_{1})+b(\mathbf{v}_{3}-\mathbf{v
}_{1})\right]^2 .
\end{eqnarray}
The components of the auxiliary functions $(\mathbf{f}_i)_k$ are selected entries of the tensor $I_T$ or zero.
$I_{T}$ can be written in terms of the
triangle vertices $\mathbf{v}_\mu$ and triangle centres $\mathbf{C}_T$ of $T$ as
\begin{eqnarray}
    I_{T} = 2\vert T\vert \left(\frac{9}{24} \mathbf{C}_{T}^2 + \frac{1}{24} \Sum{\mu=1}{3}
\mathbf{v}_{\mu}^2 \right).
\end{eqnarray}

\begin{table}
\centering
{\renewcommand\arraystretch{1.5}
\begin{tabular}{cccc|cccc}
\multicolumn{4}{c|}{$W_0^{1,0}$}&\multicolumn{4}{c}{$W_0^{2,0}$}\\\hline
$i$   & $\mathbf{ f}_i$   & $k$ &&& $i,j$ & $\mathbf {f}_{ij}$& $k$ \\\hline
$x$   & $(0,xy,0)^t$      & $y$ &&& $x,x$  & $(0,0,xxz)^t$    & $z$ \\
$y$   & $(0,0,yz)^t$      & $z$ &&& $y,y$  & $(0,0,yyz)^t$    & $z$ \\
$z$   & $(xz,0,0)^t$      & $x$ &&& $z,z$  & $(0,zzy,0)^t$    & $y$ \\
&&&&&$x,y$  & $(0,0,xyz)^t$    & $z$ \\
&&&&&$x,z$  & $(0,xyz,0)^t$    & $y$ \\
&&&&&$y,z$  & $(xyz,0,0)^t$    & $x$
\end{tabular}}
\caption{Auxiliary functions used to convert the volume integrals of
$W_0^{1,0}$ and $W_0^{2,0}$ into surface integrals.}
\label{tab:w0utilityfunctions}
\end{table}

The remaining integrals
$W_\nu^{1,0}$ with $\nu=1,2,3$ are evaluated similarly to the integrals $W_\nu$
$W_\nu^{2,0}$ (see below). The integrals $W_\nu^{0,1}$ ($\nu=1,2,3$) involving
surface
normals vanish for arbitrary bodies (with closed bounding surfaces).
\subsection{Volume integral $W_0^{2,0}$}\label{sec:w020deriv}
The volume integral $W_0^{2,0}(P)$ can be computed in a
similar way as $W_0^{1,0}(P)$.  Using
\begin{eqnarray}\label{eq:JTtensor}
    J_{T} &= 2\vert T\vert \Int 01 \d a \Int 0{1-a}\,\d b
\left[\mathbf{v}_{1}+a(\mathbf{v}_{2}-\mathbf{v}_{1})+b(\mathbf{v}_{3}-\mathbf{v
}_{1})\right]^3,
\end{eqnarray}
the components $ij$ of the tensor may be expressed as
\begin{eqnarray}
    \left(W_0^{2,0} (K) \right)_{ij} = 
        \Sum{T\in\mathcal{F}_2}{}   (J_{T})_{ijk}
(\mathbf{n}_{T})_{k}.
\end{eqnarray}
Again, the index $k$ is not a summation index but rather the index specified in
table \ref{tab:w0utilityfunctions}.
This derivation applies equally to convex and non-convex polytopes $P$.

\subsection{Surface integrals $W_1^{0,2}$ and $W_2^{0,2}$}

The computation of $W_1^{0,2}$ results in a simple sum of integrals over
triangular facets, resulting in the formulae in
Tab.~\ref{Tab:MinkowskiValuationsForTriangulations}, both for convex and
non-convex bodies.\\

The tensor $W_2^{0,2}$ is calculated by a parallel body construction, first
demonstrated for convex bodies. Consider a convex polytope $P$, and the
corresponding parallel body $P_\epsilon$. The integral over the parallel surface
is 
split up into integrals over flat facets, cylindrical patches and spherical
patches. Out of these only the cylindrical edge segments contribute, for the
same reasons as for the scalar measure $W_2$. The remaining contribution is
calculated for $\epsilon\rightarrow 0$ using the following representation for
the normal vectors on the cylindrical patches. Given an edge $\mathbf{e}$ with
facet normals $\mathbf{n}_{T}$ and $\mathbf{n}_{T'}$ of
the adjacent triangles. One obtains, also representing a special case of
eq.~(\ref{Tensorforpolytopes}),

\begin{equation}
W_2^{0,2}(K) = \frac{1}{12}\sum_{\mathbf{e}\in \mathcal{F}_1}
\vert\mathbf{e}\vert\Int{-\alpha_\mathbf{e}/2}{\alpha_\mathbf{e}/2}\mathbf{n}^2
\d\vartheta.
\label{eq:w202}
\end{equation}
To compute the integral on the right hand side we define the orthogonal unit vectors
 $\hat{\mathbf{e}}=\mathbf{e}/\vert\mathbf{e}\vert$, $\ddot{\mathbf{n}} =
(\mathbf{n}_{\mathbf{e}_T}+\mathbf{n}_{\mathbf{e}_{T'}})/\vert\mathbf{n}_{
\mathbf{e}_T}+\mathbf{n}_{\mathbf{e}_{T'}}\vert$ and
$\dot{\mathbf{n}}=\hat{\mathbf{e}}\times\ddot{\mathbf{n}}$. For a given edge,
$\mathbf{n}(\vartheta)$ can be written as
$\mathbf{n}=\cos\vartheta\,\ddot{\mathbf{n}}+\sin\vartheta \,\dot{\mathbf{n}}$. 
In this basis, the integral over $\mathbf{n}^2$ evaluates to
$(1/2)\left((\alpha_\mathbf{e}+\sin\alpha_\mathbf{e})\ddot{\mathbf{n}}
^2+(\alpha_\mathbf{e}-\sin\alpha_\mathbf{e})\dot{\mathbf{n}}^2\right)$,
see~Fig.~\ref{Fig:NotationTriangulation}. This yields the formula in
Tab.~\ref{Tab:MinkowskiValuationsForTriangulations}. The validity of this
formula for non-convex 
bodies follows from the same additivity arguments as for $W_2$.

\subsection{Curvature-weighted surface integrals $W_2^{2,0}$ and $W_3^{2,0}$}

The mean and Gaussian curvature weighted surface integrals $W_2^{2,0}$ and
$W_3^{2,0}$ over position vectors can be evaluated as the limit $\epsilon
\rightarrow 0$ of the parallel body construction, for convex bodies. The
validity for non-convex shapes follows from the analogous construction as for
$W_2$ and $W_3$ (see tab.~\ref{Tab:MinkowskiValuationsForTriangulations}).

\subsection{\label{subsec:open-bodies}Open bodies, labelled domains and Minkowski maps}

\begin{figure}[t]
\hfill
\begin{minipage}{0.85\textwidth}
  \begin{minipage}{0.49\textwidth}
    \fixsubfiglabel{a}\\
    \includegraphics[width=1\textwidth]{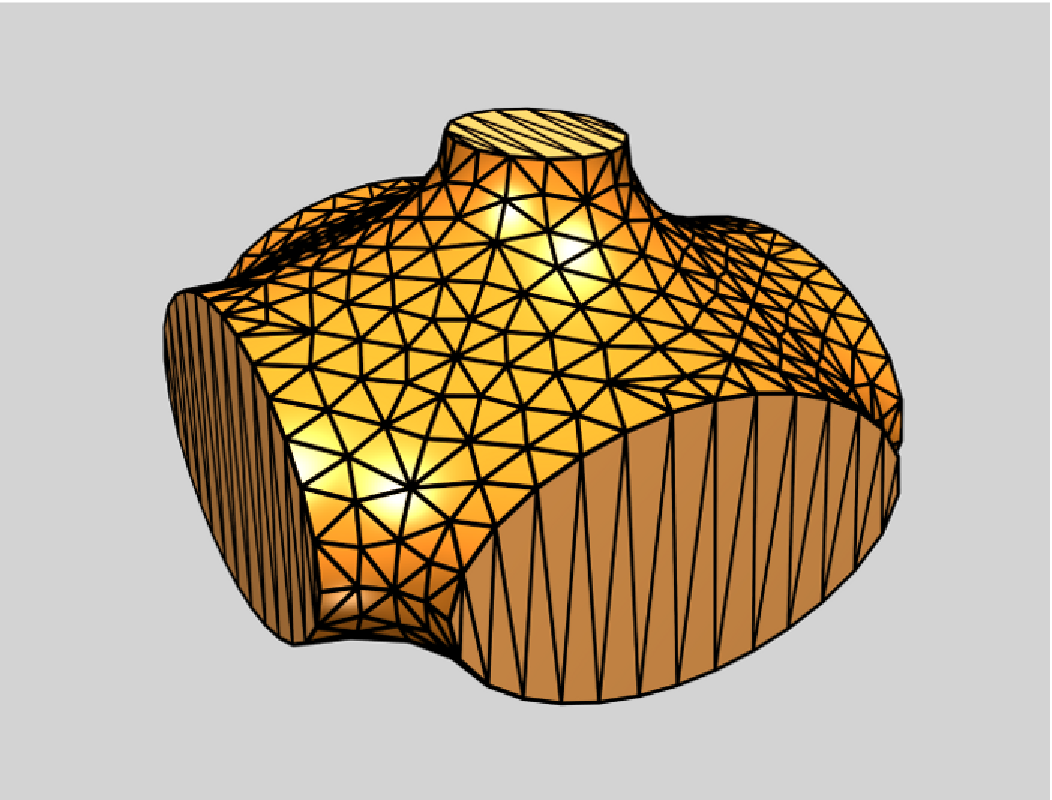}
  \end{minipage}
  \hfill
  \begin{minipage}{0.49\textwidth}
    \fixsubfiglabel{b}\\
    \includegraphics[width=1\textwidth]{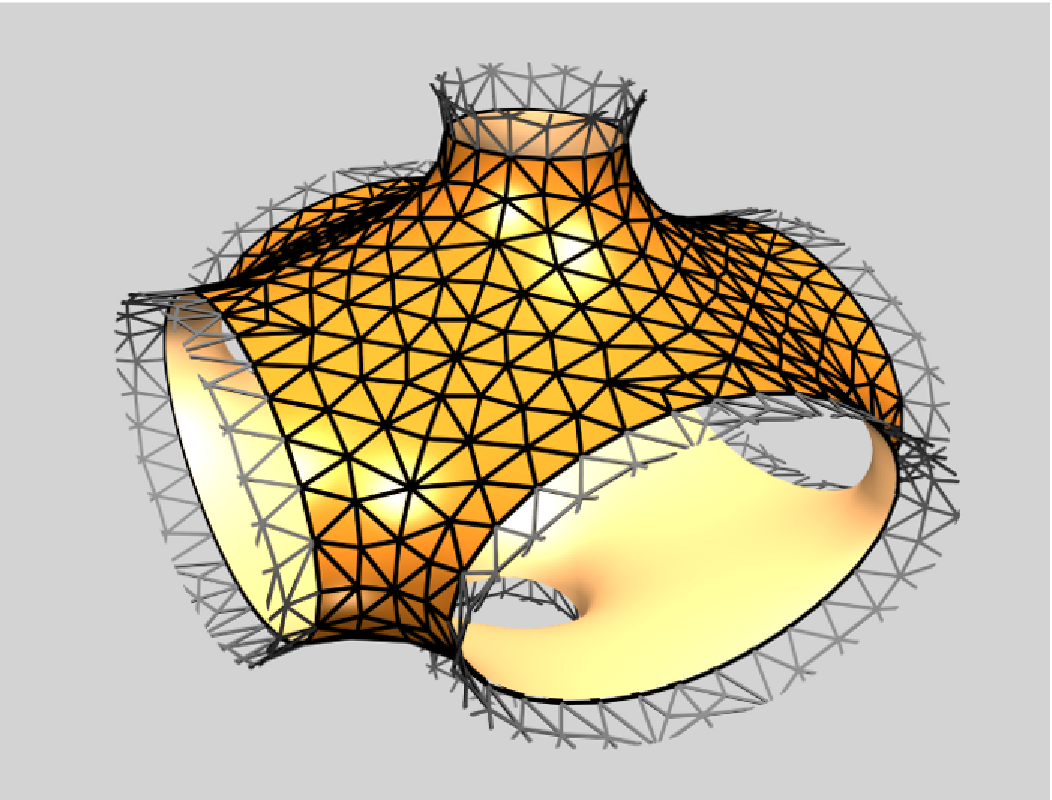}
  \end{minipage}
\end{minipage}
\caption{\fixsubfiglabel{a} A compact (but not convex) body $K$ that corresponds to a
translational unit cell of a periodic body and a triangulation of it. Its bounding surface
consists in the translational unit as well as the flat ``end caps''
that seal it. \fixsubfiglabel{b} A surface portion representing a translational unit and its triangulation.
The body $K$ is the volume to one side of the
surface and forms a connected periodic body. The surface 
and its triangulation extend beyond the translational unit cell indicating that
the surface (and the triangulation) is periodic.}
\label{fig:illustration-sealed-vs-periodic-tuc}
\end{figure}

The analysis presented so far has been derived for compact bodies in
$\mathbbm{R}^3$ with a {\em closed} bounding
surface -- and inherits strong robustness from its integral nature. For some
analyses, the requirement of closed bodies
 is too stringent. For example, experimental data sets of percolating or
periodic structures, both of which extend
infinitely through space, always represent finite subsets of the structure with
components that traverse the data set
boundaries. Similarly, an analysis of a periodic model may be restricted to a
translational unit cell, see
Fig.~\ref{fig:illustration-sealed-vs-periodic-tuc}. Furthermore, a \emph{local}
MT analysis, termed a {\em
Minkowski map}
\cite{RehseMeckeMagerle:2008,SchroederTurkKapferBreidenbachBeisbartMecke:2009},
can be useful to quantify variations throughout the sample.
For a Minkowski map, a grid is superposed on the body $K$, and the MT are computed
separately for each grid domain $L$. Such Minkowski maps can be useful to
analyse spatial heterogeneity of anisotropy or
orientation at the length scale given by the size of $L$. In these situations,
the MT are computed for the
subset $L\cap \partial K$ of the whole bounding surface that is contained in a
box $L$. In general, $L\cap\partial K$ is not a closed surface even if $\partial
K$ is.

It is evidently possible to take the subset $K\cap L$ of $K$ and consider
$\partial(K\cap L)$ as the bounding surface. However this
introduces bounding surface patches (e.g.~solid/void interfaces if $K$ is a
porous medium) that are not part of the bounding surface $\partial K$ of $K$.
For physical analyses one may want to avoid such {\em boundary effects},
i.e.~not consider the contributions of these additional bounding surface
patches. This motivates the introduction of MF and MT for open bodies,
i.e.~bodies without a closed bounding surface (see
Fig.~\ref{fig:illustration-sealed-vs-periodic-tuc}).\\

\begin{figure}
  \hfill
  \includegraphics[width=0.85\textwidth]{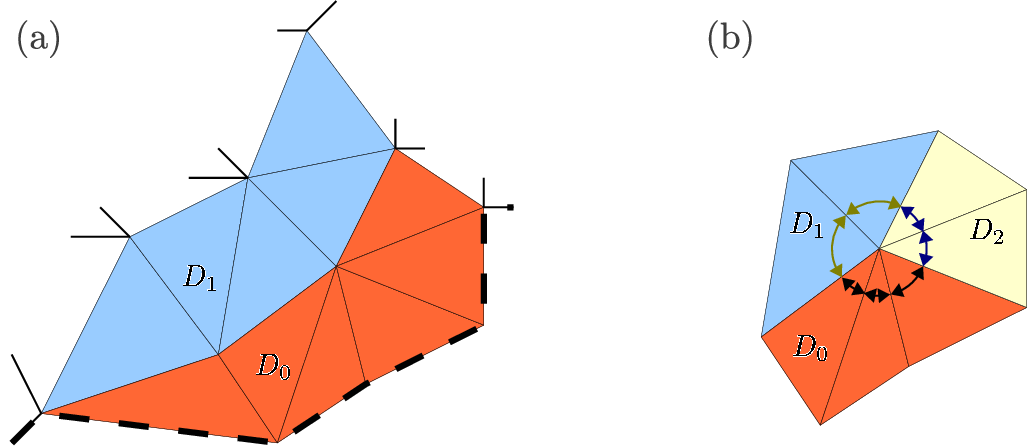}
\caption{\label{Fig:Labels}
\fixsubfiglabel{a} A triangulated surface may be decomposed into several domains by
assigning a domain label to each triangle; also open boundaries are possible,
i.e.~triangle edges without adjacent triangles (thick dashed line). \fixsubfiglabel{b} Label
factors at vertices which are adjacent to triangles of more than one domain. The
Gaussian curvature contribution is weighted with the interior angles
belonging to each domain.
}
\end{figure}

In lieu of an attempt to define MF and MT for open bodies, we define a domain-wise analysis
of MF and MT. Consider a decomposition of the surface $\partial P$ of a triangulated body $P$ into $m$ 
 domains, or patches, $D_\sigma$ (with $\sigma=1,\dots,m$) such that
\begin{equation}\label{eq:subdomains}
 \partial P= \bigcup_{\sigma=1}^{m}D_\sigma,
\end{equation}
 and consider these domains {\em label ed} by labels $\sigma=1,\dots, m$.  Triangles are uniquely assigned to a label,
but edges and vertices of the triangulation can be shared between several domains. Specifically, the domains $D_\sigma$ could
represent a decomposition of $P$ into patches contained within the grid domains of a three-dimensional lattice (See Fig.~\ref{Fig:Labels}).

Contributions of the facets 
can be uniquely assigned to $D_\sigma$, 
but the contribution of edges and vertices on the boundaries of a patch $D_\sigma$ needs to be divided between $\sigma$ and
the label of the adjacent domain $\sigma'$
(See Fig.~\ref{Fig:Labels}).

For $W_2^{r,s}$, the contribution of the dihedral angle at an edge is equally divided between the labels of the two adjacent
triangles (Note that this is naturally taken account of by the use of oriented edges in the doubly-connected edge list,
discussed above). For $W_3^{r,s}$ the division of the contribution of the interior vertex angles to the integral Gaussian
curvature measures $W_3^{r,s}$ is less straight-forward. An intuitive way, that is also consistent with global
integration over all labelled domains, is provided by the use of {\em label factors}. 
The label factor $f_{D_\sigma}(\mathbf{v})$ of domain $D_\sigma$ at vertex $\mathbf{v}$ is defined
as
\begin{equation}
f_{D_\sigma}(\mathbf{v}):=\frac{
\Sum{T\in \mathcal{F}_2(\mathbf{v})\cap\mathcal{F}_2
(D_\sigma)}{}\phi_{T}^{\mathbf{v}}
}{ \Sum{T\in \mathcal{F}_2(\mathbf{v}) }{}\phi_{T}^{\mathbf{v}}},
\end{equation}
where $\mathcal{F}_2(D_\sigma)$ is the set of all triangles labelled with label
$\sigma$. Hence $f_{D_\sigma}(\mathbf{v})$ is the sum of angles at $\mathbf{v}$ of those
triangles adjacent to $\mathbf{v}$ and are labelled $\sigma$ divided by sum of
these angles of all adjacent triangles.
It is easy to see, that
\begin{equation}
w_3(\mathbf{v})=\sum_{\sigma=0}^{m} f_{D_\sigma}(\mathbf{v}) w_3(\mathbf{v}),
\end{equation}
for a vertex with $m$ adjacent labels and $W_3(P)=\sum_{\mathbf{v}\in\mathcal{F}_0}w_3(\mathbf{v})$.

For the volume tensor $W_0^{2,0}$ a label-wise analysis is only well-defined if
the body $K$ is subdivided (and not only the bounding surface $\partial K$).

{
\subsection{Implementation details and ``karambola'' software package}

A fully functional implementation of the algorithms represented in sections \ref{subsec:volume} to \ref{subsec:open-bodies}  is provided as online supplementary material to this article, and also made available through the internet at {\tt www.theorie1.physik.fau.de/karambola}, under a GNU General Public License.

The implementation is a straightforward realisation of the formulae in the rightmost column of Tab.~\ref{Tab:MinkowskiValuationsForTriangulations} into ANSI-C code. A simple data structure is used to store a triangulation of a surface, as a set of points and a list of facets (specifically triangles); the data structure allows to iterate over all vertices, edges or facets by simple loops, for example over all edges ``{\tt for $e$ in $\mathcal{F}_1$}'', and it allows the extraction of neighbours, for example of all (one or two) triangles that are adjacent to a given edge $e$. With that data structure in place, the sums in Tab.~\ref{Tab:MinkowskiValuationsForTriangulations} simply become {\tt for} loops, that are all linear in the number of edges, facets or vertices. 
}

\section{Anisotropy Measures}
\label{sec:anisotropy-measures}

Based on MT, robust measures of
anisotropy can be defined that are sufficiently sensitive to capture subtle
anisotropy effects and that are applicable to a wide range of microstructures. The
usefulness and versatility of this approach is demonstrated by two examples
representing different types of structures -- a cellular partition and
a network structure.

A rank-2 tensor is defined to be \emph{isotropic} if and only if
it is proportional to the unit tensor $Q$,
i.e.~its eigenvalues are all equal. Deviations from isotropy are measured by the
anisotropy index $\beta_\nu^{r,s}$, which is the ratio of extremal eigenvalues
of the tensor $W_\nu^{r,s}$.
For example, let $\xi_\mu$ ($\vert\xi_1\vert\leq\vert\xi_2\vert\leq \vert\xi_3\vert$) be the
eigenvalues of $W_1^{0,2}$, then the anisotropy index is
\begin{equation}
    \beta_1^{0,2} :=
    \left|\frac{\xi_{1}}{\xi_{3}}\right|\in [0,1].
    \label{eq:anisotropy-measure-W102}
\end{equation}
By definition, this quantity is dimensionless, continuous and rotation invariant.
The value of 1 indicates perfect isotropy, and smaller values indicate
anisotropy.
For anisotropic bodies, it is sometimes also useful to consider
$\gamma_1^{0,2}=|\xi_2/\xi_3|$.

These quantities can be easily interpreted for the translation invariant
tensors $W_1^{0,2}$ and $W_2^{0,2}$. 
We can write $W_1^{0,2}$  equivalently to eq.~(\ref{eq:defMinktens}) 
as the second moment of the distribution of normal vectors (with the density
$\rho_1(\mathbf{n})$)
on the unit sphere $\mathbbm{S}^{2}$ as
\begin{equation}
W_1^{0,2}(K) = \frac{1}{3}\int_{\mathbbm{S^2}} \rho_1(\mathbf{n})\;
\mathbf{n}\odot\mathbf{n}\,\d\Omega,
\label{eq:express-W102-by-normal-distribution}
\end{equation}
 where the $\rho_1(\mathbf{n}')=\int_{\partial K} \delta(\mathbf{n}-\mathbf{n}') \d A$. That is,
$\rho_1$ is area-weighted density of normal vectors.  It is easy to see that an
uniform distribution on $\mathbbm{S}^2$ is equivalent to an isotropic tensor.

For example, if $K$ is a sphere, then $\rho_1(\mathbf{n})$ is constant and
$\beta_1^{0,2}=1$ as expected. For the rectangular box $[0,a_x] \times [0,a_y]
\times [0,a_z]$, the function $\rho_1(\mathbf{n})$ is concentrated at delta
peaks,
\begin{equation}
\rho_1 (\mathbf n) = a_x a_y a_z   \sum_{i = x,y,z}
 \frac{\delta (\mathbf{e}_i - \mathbf{n})  + \delta (\mathbf{e}_i + \mathbf{n})}{ a_i}
\end{equation}
The resulting anisotropy measure is $a_z/a_x$ for $a_x\ge a_y \ge a_z$.

It is instructive to express the second translation-invariant MT
$W_2^{0,2}$ by a distribution of 
normals and curvatures. The density 
\begin{equation}
\rho_2(\mathbf{n}', G_2') = \int_{\partial K} \delta(\mathbf{n}-\mathbf{n}')
\delta(G_2-G_2') \mathrm{d} A
\label{eq:distribution-curvatures-nromals}
\end{equation}
gives the sum of the area of all surface patches that have normal direction
$\mathbf{n}'$ and mean curvature $G_2'$. 

\begin{equation}
W_2^{0,2}(K) = \frac{1}{3}\int_{-\infty}^\infty  G_2
\int_{\mathbbm{S^2}}
  \rho_2(\mathbf{n},G_2) \, \mathbf{n}\odot\mathbf{n} \,\mathrm{d} \Omega\,\mathrm{d}G_2.
\label{eq:express-W202-by-normal-distribution}
\end{equation}
If the function $\rho_2$ can be written as a product
$\rho_2(\mathbf{n},G_2)=\tilde \rho_2(G_2)\,\rho_1(\mathbf{n})$, the anisotropy
characteristics $\beta_1^{0,2}$ and $\beta_2^{0,2}$ defined as the ratio of the
smallest to the largest eigenvalue of $W_2^{0,2}$ are identical. In this sense,
$\beta_2^{0,2}$ provides a higher order anisotropy measure that quantifies
anisotropy of the curvature distribution.\\

\subsection{Alignment of Actin Biopolymer networks under shear}
\label{Sec:actin-filamin-alignment}

\begin{figure}[t]
    \setlength{\unitlength}{0.75\textwidth}
    \hfill
    \begin{minipage}{\unitlength} 
    \begin{picture}(1,0.9)

\put(0.0,0.0){\includegraphics[width=\textwidth]{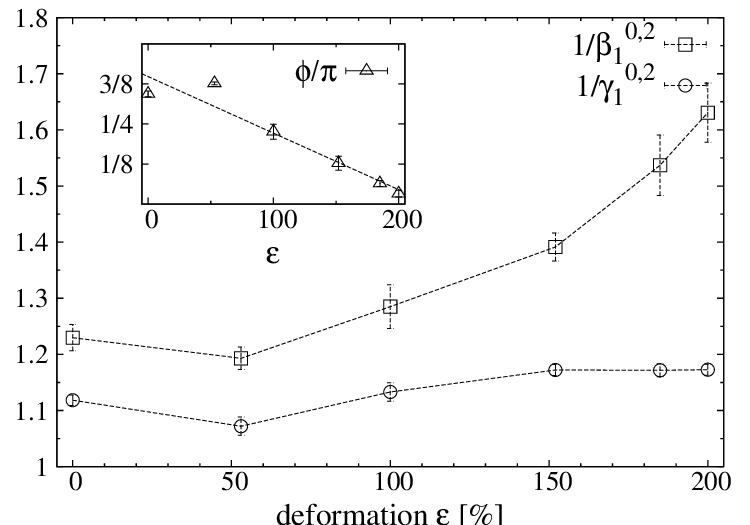}}

\put(0.0,0.7){\includegraphics[height=0.17\textwidth]{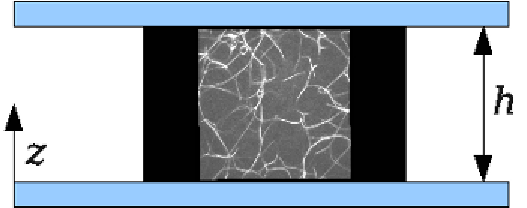}}

\put(0.47,0.7){\includegraphics[height=0.17\textwidth]{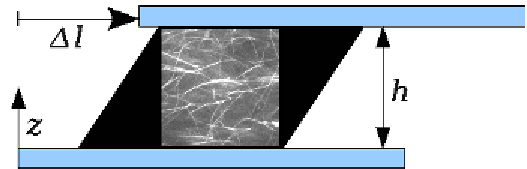}}
    \end{picture}
  \end{minipage}
  \caption{Anisotropy measures $1/\beta_1^{0,2}$ and $1/\gamma_1^{0,2}$ of actin
network as function of shear. The ratio of largest to the smallest eigenvalue
$1/\beta_1^{0,2}$ grows with increasing shear indicating that the fibbers become
increasingly aligned. The ratio of the largest to the intermediate eigenvalue
$1/\gamma_1^{0,2}$ remains close to unity as expected since the fibres are
essentially one-dimensional lines, inflated to approximately cylindrical tubes.
The insert shows the alignment angle $\phi$ of the eigenvector corresponding to
the minimal eigenvalue $\xi_1$ of $W_1^{0,2}$ and the direction of
applied shear; its decay to 0 indicates that the network aligns with the shear
direction when large shear is applied. The error bars in both plots are the
standard deviation of the distribution of the quantities when analysed for
different segmentation parameter between $0.95<\Lambda<0.99$. The illustrations
above sketch the experimental setup with confocal microscopy images
corresponding to a fibre network at shear $\epsilon:=\Delta l/h = 0$ and
$\epsilon=2$ (the small confocal microscopy images are taken from
\cite{LiuKaszaKoenderinkVaderBroederszMacKintoshWeitz:2008}).}
  \label{fig:actin-filamin-shear}
\end{figure}

Biopolymer networks made of actin or collagen fibres are important structural
elements in biological tissue that act as a scaffolds and provide stiffness and
mechanical stability \cite{GardelShinMacKintoshMahadevanMatsudairaWeitz:2004,CukiermanPankovStevensYamada:2001, PedersenSwartz:2005}.
Of current interest is the
relationship between fibre arrangement and alignment on the one hand side and
elastic or visco-elastic properties on the other. This relationship can be
probed by shear-experiments with confocal microscopy providing real-space
structural data \cite{LiuKaszaKoenderinkVaderBroederszMacKintoshWeitz:2008}. We now demonstrate that the degree of alignment and of
structural anisotropy of the fibre network is well-captured by a MT analysis.

The data sets analysed here represent Actin fibre networks reconstituted from
rabbit actin biopolymer networks with actin concentration of $1.2\mathrm{mg/ml}$
cross-linked with filamin A. These are imaged using confocal microscopy for
different shear deformations, see the explanation in
Fig.~\ref{fig:actin-filamin-shear}. The data sets are the same as those analysed
in \cite{LiuKaszaKoenderinkVaderBroederszMacKintoshWeitz:2008}. The gray-scale
data set is converted into a binary data set with 1 corresponding to actin and 0
corresponding to the surrounding fluid by standard threshold segmentation with
threshold $I_c$ \footnote{The threshold $I_c$ is chosen such that only the
brightest and
hence thickest fibres are retained. For a given segmentation threshold
$I_c$ the integrated intensity of all voxels of the fluid phase is
$\Lambda=(\sum I(p))^{-1}\sum^{*} I(p)$ where $I(p)$ is the intensity of voxel
$p$ in the original intensity data set, $\sum$ the sum over all voxels of the
data set and $\sum^{*}$ the sum over all voxels of the fluid phase, i.e.~those
voxels that are set to $0$ by the segmentation process. The values of $I_c$
chosen here correspond to $0.95<\Lambda<0.99$.}. The medial axis of the 1 phase
is computed using distance-ordered homotopic thinning 
\cite{MaSonka:1996,LeeKashyapChu:1994} and is used as the one-voxel thick line
representation of the actin fibre network. Conversion to a triangulated representation is
obtained by using the Marching Cubes algorithm \cite{LorensenCline:1987}. For more
details of the analysis of biopolymers see
ref.~\cite{MickelMuensterJawerthVaderWeitzSheppardMeckeFabrySchroederTurk:2008}.

Typically only a subset, or observation window, of the
structure is available for analysis.
Therefore we assume that the network is homogeneous and a sufficiently large
but finite subset is accessible
which is assumed to represent the entire sample.
The derived measures $\beta_1^{0,2}$ quantify the
intrinsic anisotropy, i.e.~their values do not depend on the size, aspect
ratio or position of the observation window (for sufficiently large windows).

Figure~\ref{fig:actin-filamin-shear} shows $1/\beta_1^{0,2}$ and $1/\gamma_1^{0,2}$ evaluated on the whole network (that consists essentially in a
single component; only the outer boundary layers of the confocal data are clipped).
It shows that the distribution of normal directions of the
fibre bounding surface becomes less isotropic with increasing shear, indicating
alignment of the fibres. The angle between the eigenvector to the minimal
eigenvalue $\xi_1$ (corresponding approximately to the dominant tangent direction) and
the direction of shear decreases to 0, indicating the alignment of the fibres
with the direction of shear. This is commensurate with the results published in
\cite{LiuKaszaKoenderinkVaderBroederszMacKintoshWeitz:2008},
that extracted a distribution of tangent directions and used these to quantify
alignment.

The eigenvalue ratios of the translation-covariant tensors $W_\nu^{2,0}$ (and of
the tensor of inertia) capture different aspects of the anisotropy of a shape
compared to the translation invariant tensors $W_\nu^{0,2}$, see also section
\ref{subsec:Mink-Tenso-of-conv-polyh}. However, usefulness of the translation-covariant tensors depends on
whether or not a natural definition of the origin is available for the system.
For example, for the analysis of cellular shapes one may choose the centre of mass
$W_0^{1,0}/W_0$ or the corresponding curvature centroid $W_\nu^{1,0}/W_\nu$ as
the origin. Especially for percolating or periodic bodies, for which the
analysis
is always restricted to a finite window of observation, the choice of origin is
 often not naturally determined. An additional problem for such
structures is that the measures $\beta_\nu^{2,0}$ derived from translation
covariant tensors $W_\nu^{2,0}$, as opposed to the translation-invariant
measures $\beta_\nu^{0,2}$, crucially depend on the shape and size of the window
of observation.

The analysis of alignment of biopolymer networks illustrates the potential of
the MT approach for structure characterisation of cellular and
porous materials, and demonstrates its applicability to voxelised experimental
data. The MT approach can shed light on systems with a similar
spatial structure that exhibit subtle anisotropy effects, such as fibrous
biological materials \cite{Breidenbach:2007}, porous
materials\cite{SchroederTurk:2011b}, and metal foams \cite{SaadatfarEtAl:2011}.

\subsection{Anisotropy of free-volume cells of random bead
Packs}\label{Sec:free-volume-cells-tomasos-bead-packs}

Granular media represent a system where geometry plays an essential role in
determining its physical properties, such as flow or packing properties. The
geometric structure to be characterised is substantially different from the
above example. It consists in an assembly of (disjoint) grains that have, at most, mutual point contacts.

A commonly used way to associate each grain with {\em its} corresponding region of space is the construction of a Voronoi diagram. The distributions of volumes of the Voronoi cells of disordered jammed sphere
packs with packing fractions from 0.55 (random loose packing) to around 0.64
(random close packing, RCP) have attracted interest for the study of granular
systems \cite{Aste:2006}, motivated by a possible statistical mechanics
description of granular systems \cite{MehtaEdwards:1989,Zamponi:2008}.
For instance, Aste {\em et al.}~\cite{Aste:2006} used the volume distribution of Voronoi cells to estimate
configurational entropy in static packings, and Zhao {\em et al.}~\cite{ZhaoSidleSwinneySchroeter:2012} to quantify spatial correlations in disc packings.

Here we illustrate how Minkowski tensors can be used to characterise the shape, rather than simply the volume, of the grains' Voronoi cells, in the spirit of reference \cite{SchroederTurkMickelSchroeterDelaneySaadatfarSendenMeckeAste:2010}.

\begin{figure}[t]
\hfill
\begin{minipage}{0.8\textwidth}
  \newlength{\imageheight}
  \setlength{\imageheight}{0.4\textwidth}
  \begin{minipage}{1.02\imageheight}
    \fixsubfiglabel{a}\\
    \includegraphics[height=\textwidth]{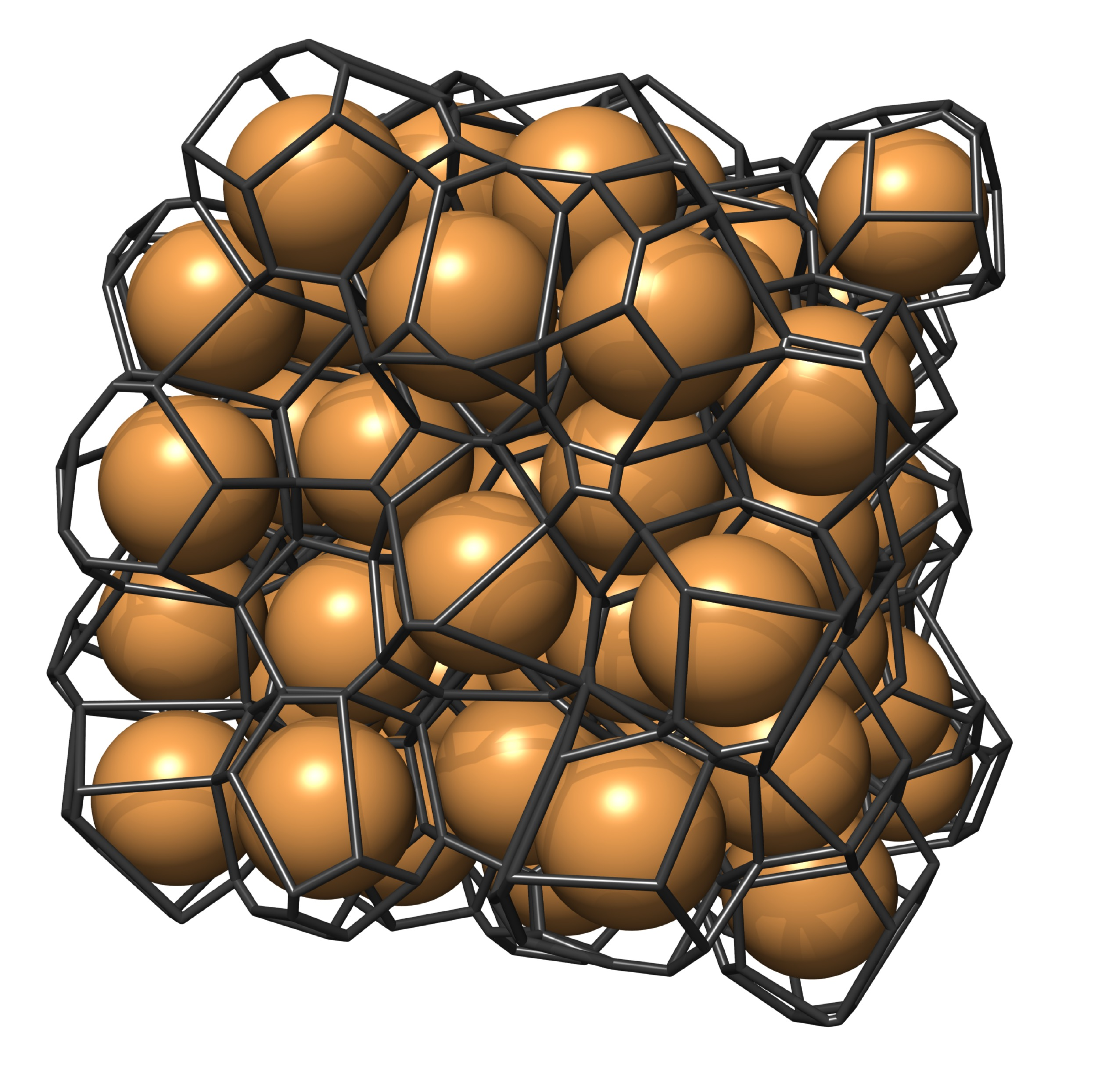}
  \end{minipage}
  \begin{minipage}{1.2\imageheight}
    \fixsubfiglabel{b}\\
    \includegraphics[height=\imageheight]{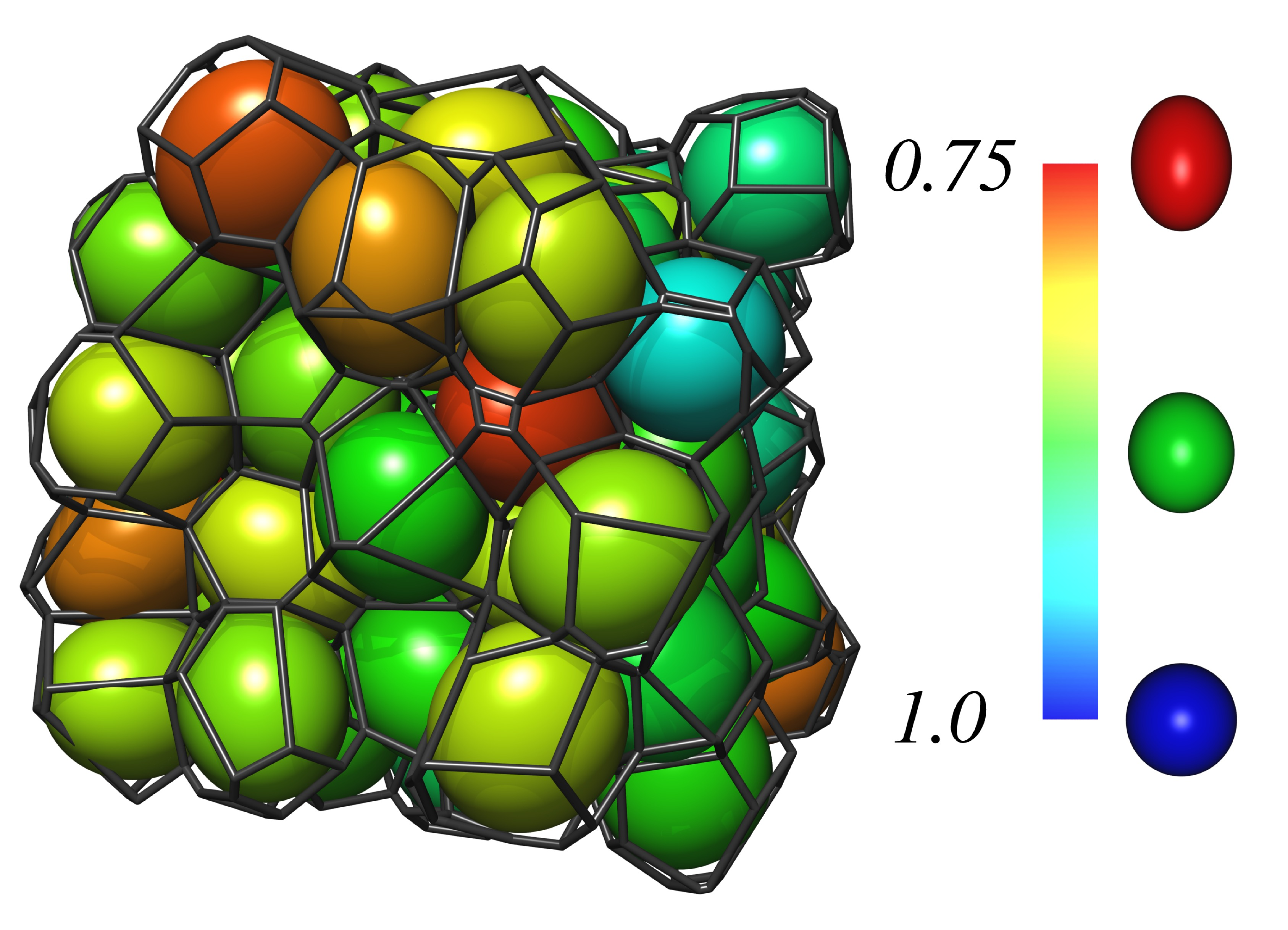}
  \end{minipage}
\end{minipage}
\caption{\label{fig:Ellipsoids} \fixsubfiglabel{a} Voronoi diagram of jammed
monodisperse sphere pack. \fixsubfiglabel{b} The same Voronoi diagram, but spheres
are replaced by ellipsoids with half-axes $a_1=a_2<a_3$, aligned along the
eigendirections of $W_0^{2,0}$. Colours represent the ratio of the shortest and
longest axis of the ellipsoid. See also Fig.~\ref{fig:eig-val-ratio-ellipsoid} in the appendix for the relation of the half axes ratios $a_1/a_3$ to the MT anisotropy measure $\beta_0^{2,0}$ for ellipsoidal particles. (figure reproduced from ref.~\cite{SchroederTurkMickelSchroeterDelaneySaadatfarSendenMeckeAste:2010}).
}
\end{figure}

Figure \ref{fig:Ellipsoids} shows a subset of an experimental data set of static disordered monodisperse jammed spheres with
packing fraction 0.58 on the left panel (details see ref.~\cite{SchroederTurkMickelSchroeterDelaneySaadatfarSendenMeckeAste:2010})
The wire-frame illustrates the Voronoi diagram of these spheres \footnote{The Voronoi cells of the sphere centres are computed using the program \emph{qhull} \cite{BarberDobkinHuhdanpaa:1996}. For a point $\mathbf{p}$ of a set $\mathcal{P}$ of points, its Voronoi cell is the convex polytope that contains all points of $\mathbbm{R}^3$ closer to $\mathbf{p}$ than to any other point in $\mathcal{P}$ \cite{OkabeBootsSugiharaChiuOkabe:2000}.}.  
On the right hand side, spheres were replaced by ellipsoids that have the same eigenvalue ratio of the Minkowski tensor $W_0^{2,0}$ as their Voronoi cells, $W_0^{2,0}(\mathrm{ellipsoid})=W_0^{2,0}(\mathrm{Voronoi\ cell})$, implying in particular the same value of the 
anisotropy measure $\beta_0^{2,0}$. Their half-axes are aligned with the eigendirections of
the tensor $W_0^{2,0}$, evaluated for the corresponding Voronoi cell. The use of MT, and their
eigenvalues and eigendirections, hence provides an efficient way of `fitting'
ellipsoids to given convex cells and hence a means to quantify their anisotropy or elongation. This in turn provides sensitive tools to quantify shape and structure of both amorphous and ordered particulate systems and packings.

Quantitative analyses of the Voronoi cell shapes of disordered sphere configurations, in various phases, have been given in references \cite{SchroederTurkMickelSchroeterDelaneySaadatfarSendenMeckeAste:2010,KapferMickelMeckeSchroederTurk:2012,KapferMickelSchallerSpannerGollNogawaItoMeckeSchroederTurk:2010,MickelKapferSchroederTurkMecke:2013}, illustrating the breadth of the Minkowski tensor approach and in particular its usefulness to capture the onset of crystallisation. An application to ordered sphere packings has been given in preliminary form in the conference proceeding \cite{SchroederTurkSchieleinKapferSchallerDelaneySendenSaadatfarAsteMecke:2013}.

\begin{figure}[t]
\hfill
\begin{minipage}{0.8\textwidth}
    \includegraphics[width=\textwidth]{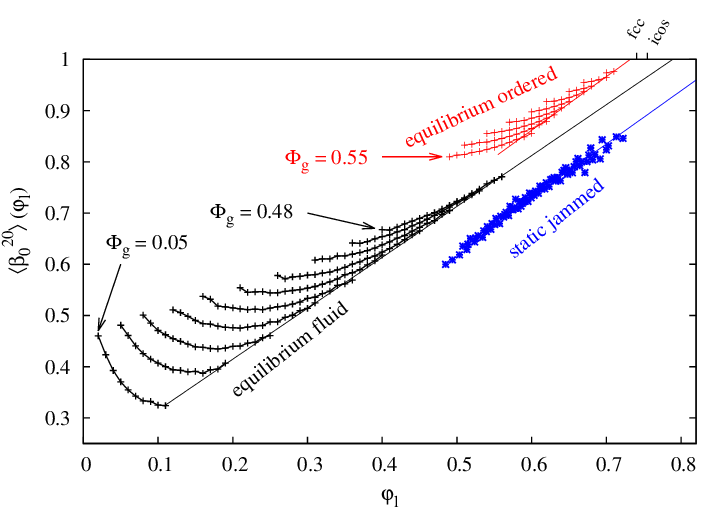}
\end{minipage}
\caption{\label{fig:local-anisotropy-local-pf} Voronoi cell anisotropies quantified by $\beta_0^{2,0}$ as function of local packing fraction $\phi_l$, for equilibrium hard spheres in the fluid phase and in the ordered phase and for jammed static disordered sphere packs. The data representing static jammed spheres represent 6 distinct sphere configurations (including tomographic images and simulations with and without friction and gravity \cite{StillingerLubachevsky:1993,JerkinsSchroeterSwinneySendenSaadatfarAste:2008,AsteSaadatfarSenden:2006}) each with a different global packing fraction $0.585\le \phi_g \le 0.64$; these are the same used in Fig.~6 of \cite{SchroederTurkMickelSchroeterDelaneySaadatfarSendenMeckeAste:2010}. The equilibrium hard sphere data is obtained from Monte Carlo simulations, and comprise 4000 to 16000 spheres, see ref.~\cite{KapferMickelSchallerSpannerGollNogawaItoMeckeSchroederTurk:2010} for details. ``fcc'' marks the packing fraction of the densest crystallographic sphere pack $\phi_g=\phi_l=\pi/\sqrt{18}\approx 0.7404$ and ``ico'' the densest possible local configuration $\phi_\mathbf{ico}=(25+11\sqrt{5})^{3/2}\pi/[15\sqrt{10}(15+7\sqrt{5})]\approx 0.7547$. The straight lines represent guides-to-the-eye only for what may be the common asymptotic behaviour of the sets of curves of each phase. Note in particular that the data for static jammed spheres appears to collapse to a single curve, independent on global packing fraction and packing protocol, in contrast to the equilibrium systems. (A script to generate this plot is added as online supplementary material, in the {\tt demo} subfolder of the {\tt karambola} Minkowski tensor program.)}
\end{figure}

The scope of Minkowski tensors for the analysis of granular material is however not restricted to the detection of local crystalline domains. Rather, as the following analysis illustrates, these shape measures also allow for a quantitative description of the local structure in {\em amorphous} assemblies, discriminating sharply between different types of amorphous geometries. 

Figure \ref{fig:local-anisotropy-local-pf} shows a study where the simple anisotropy measure $\beta_0^{2,0}$ derived from the Minkowski tensors reveals a distinct difference between the different phases that hard sphere systems can adopt: the plot shows the typical shape, quantified by $\beta_0^{2,0}$, of a Voronoi cell of a given volume, expressed as the local packing fraction $\phi_l := W_1(\mathrm{grain})/W_1(\mathrm{Voronoi\ cell})$. The plot contains data for equilibrium hard sphere systems in the fluid and in the ordered phase (the same data as used in reference \cite{KapferMickelSchallerSpannerGollNogawaItoMeckeSchroederTurk:2010}) and of jammed static sphere packs (the same data sets mentioned above and used in Fig.~6 of ref.~\cite{SchroederTurkMickelSchroeterDelaneySaadatfarSendenMeckeAste:2010}). For each sphere configuration, the Voronoi cells of all particles are computed and their {\em local packing fractions} $\phi_l$. Then all cells are classified by their value of $\phi_l$, and the average cell shape $\langle\beta_0^{2,0}\rangle(\phi_l)$ is determined by averaging $\beta_0^{2,0}$ over all cells of a data set that have, up to a discretisation interval $\Delta \phi_l$, the same value of $\phi_l$. The curves $\langle \beta_0^{2,0}\rangle(\phi_l)$ describe Voronoi cell anisotropy as function of the local packing fraction $\phi_l$.

Figure \ref{fig:local-anisotropy-local-pf} elucidates how Minkowski tensors can help discern some of the morphological differences between the different hard sphere phases: first, this analysis of the average cell anisotropy clearly discerns the distinct geometries between the static jammed packings and the (also disordered) equilibrium fluid configurations and, more expectantly, the differences to the equilibrium ordered phase. Computing $\langle \beta_0^{2,0}(\phi_l)\rangle$ provides a signature of the origin of the data sets, clearly discerning the structure of equilibrium hard spheres and of the static jammed packings.

Second, the functional form $\langle \beta_0^{2,0}\rangle(\phi_l)$ is the same for six jammed static configurations shown here. These data sets comprise different global packing fractions and preparation protocols, including tomography data of dry acrylic beads \cite{AsteDiMatteoSaadatfarSendenSchroeterSwinney:2007} and glass beads settled against a fluid current \cite{AsteDiMatteoSaadatfarSendenSchroeterSwinney:2007}, as well as Lubachevsky-Stilinger simulations of frictionless particles without gravity \cite{LubachevskyStillingerPinson:1991} and discrete element method simulations of spheres with friction and gravity \cite{DelaneyInagakiAste:2007}. This suggests a universality of the jammed static disordered spheres, in the sense of the following observation: the typical shape (quantified by $\beta_0^{2,0}$) of a Voronoi cell of a given local packing fraction $\phi_l$ is the same for all packings, regardless of protocol and of the global packing fraction $\phi_g$. Note that this is in stark contrast to the case of the equilibrium fluid or ordered phase, where particularly small cells in a globally denser packing are more isotropic with larger $\beta_0^{2,0}$ than the same size cell in a looser packing.  This result suggests that the global packing fraction of a jammed static disordered sphere packing is the result of combining typical ``building blocks'' with a given local $\phi_l$ in different proportions. This observation may go some way towards clarifying why random close packing of spherical particles appears to be largely protocol-independent. 

Figure \ref{fig:local-anisotropy-local-pf} has provided a test case where Minkowski tensors concisely discern the differences between two types of amorphous structures, in a quantitative fashion with an intuitive geometric real-space interpretation. The characterisation of amorphous cellular shapes is also
relevant in various other contexts, e.g.~for the relationship between structure
and dynamics in glass-forming liquids \cite{StarrSastryDouglasGlotzer:2002} or
for packing entropy of the hard micellar cores in supramolecular micellar
materials \cite{ZiherlKamien:2000}, and for the understanding of physical properties
of foams, including rheological \cite{KablaDebregeas:2003} and static
\cite{KraynikReineltVanSwol:2004} properties and the evolution under ageing or coarsening \cite{evans2012,SayeSethian:2013}. Minkowski tensors can be applied to these systems in an analogous fashion to the analysis of this section.

\section{Conclusion and Outlook}
\label{sec:discussion-conclusion}

This article  provides the theoretical description and explicit algorithms for the use of Minkowski tensors in spatial structure and morphology characterisation in the natural sciences. Minkowski tensors, defined on the rigorous basis of integral geometry and endowed with strong statements about completeness, additivity and continuity, are natural extensions of the scalar Minkowski functionals. Because of their tensorial nature, Minkowski tensors allow for a quantitative evaluation of {\em shape} of anisotropic and orientation-dependent morphologies, on all length scales.\\

While the most fundamental definition of both scalar and tensorial Minkowski functionals is based on measure theoretic concepts from integral geometry \cite{SchneiderWeil:2008}, an alternative but equivalent definition based on curvature-weighted surface integrals is more intuitive for the reader without a background in measure theory. This approach also lends itself directly to numerical discretisation, resulting in the fast linear-time algorithms derived in this article. These algorithms have been comprehensively described and theoretically validated here for all relevant Minkowski tensors up to rank two, but can be generalised also for higher rank. An implementation of the algorithms described here is available as supplementary material, see below.\\ 

Minkowski tensors are versatile tools to quantitatively characterise morphological aspects related to orientation, anisotropy and elongation. Applications can be conceived in diverse fields, from nanostructures in softmatter to large-scale structures e.g.~in background radiation sky maps \cite{GoeringKlattStegmannMecke:2013}.

For particulate assemblies such as granular matter or structural glasses, our previous analyses of random jammed sphere packing systems \cite{SchroederTurkMickelSchroeterDelaneySaadatfarSendenMeckeAste:2010,KapferMickelMeckeSchroederTurk:2012} have contributed to clarifying the onset of order near the random close packing transition for spherical particles. Beyond this identification of order in these amorphous packings, Minkowski tensors lend themselves to the more complex task of quantitatively evaluating the structural changes in evolving amorphous packings, that never reach a state even with partial order, see also the discussion in sect.~\ref{Sec:free-volume-cells-tomasos-bead-packs}. In this context, the relationship between the Minkowski tensors and the bond orientational order parameters, defined by Steinhardt {\em et al} \cite{SteinhardtNelsonRonchetti:1984} as the lowest-order rotation invariants of the moments of a multipole expansion of the orientational distribution of nearest-neighbour bonds, leads to new insight into this question. We have already demonstrated that a simple idea based on Minkowski tensor analysis can remedy some of the drawbacks of the bond orientational order parameters, leading to robust {\em Minkowski structure metrics} \cite{MickelKapferSchroederTurkMecke:2013}. A formal correspondence has been shown between the Minkowski tensors $W_1^{0,n}$ of rank $n=0,1,2,\dots$ and the set of bond orientational order parameters \cite{Kapfer:2011,Mickel:2011}. This provides geometric interpretation for both methods, e.g.~the fact that $\beta_1^{0,2}$ and $q_2$ represent the same morphological information, see Fig.~5 in \cite{MickelKapferSchroederTurkMecke:2013}. This motivates a more general study also for the other Minkowski tensors such as the curvature-weighted tensors $W_\nu^{0,n}$.

A second aspect that emphasises the usefulness of Minkowski tensors for particulate matter concerns the case of aspherical particles, such as tetrahedra and polyhedra, ellipsoids and super-ellipsoids, etc, all of which have received attention in recent studies \cite{DelaneyCleary:2010,HajiAkbariEtAl:2009,NeudeckerUlrichHerminghausSchroeter:2013,ManDonevStillingerSullivanRusselHeegerInatiTorquatoChaikin:2005,DonevCisseSachsVarianoStillingerConellyTorquatoChaikin:2004}. In contrast to measures based on nearest-neighbour bonds, the Minkowski analysis naturally applies to these aspherical and possibly polydisperse particles, provided the Voronoi diagram is suitably defined. For example, with all tools in place for imaging experimental ellipsoid packings \cite{SchallerNeudeckerSaadatfarDelaneyMeckeSchroederTurkSchroeter:2013} and determining their Voronoi diagrams \cite{ellipsoid_voronoi_draft:2013}, the Minkowski tensor analysis may shed light on the more complicated, possibly less universal, mechanisms involved in jamming of ellipsoidal particles.
 
These results and the Minkowski tensor analysis itself are likely to provide new insight also for other problems in particulate systems where spatial structure largely determines physical properties, such as static and rheological properties of glass-forming systems or liquids out of equilibrium.

Importantly, however, Minkowski tensor analysis is not restricted to particulate systems. To name one further example, Minkowski tensors can aid anisotropy or alignment studies of porous materials, cellular structures and other structures that consist essentially in a single connected component that percolates macroscopically, see also section \ref{Sec:actin-filamin-alignment}. In addition to the use of Minkowski tensors as robust tools to extract anisotropy and morphology measures from tomographic or confocal microscopic images \cite{SchroederTurk:2011b,SaadatfarEtAl:2011}, these measures may also be amenable to analytical treatment for some important mathematical models of porous materials. In particular for the case of the anisotropic Boolean model, it is feasible to obtain analytic expressions for the mean values of the translation-invariant Minkowski tensors $W_1^{0,2}$ and $W_2^{0,2}$ \cite{HoerrmannKlattMeckeHug:2013}. For ordered porous structures, analytic formulae for the Minkowski tensors of triply-periodic minimal surfaces have been derived from the Weierstrass parametrisation \cite{MickelSchroederTurkMecke:2012}. The feasibility of at least some analytic treatment hints at the role that the Minkowski tensors can play for an improved understanding of the spatial structure of porous materials, and in particular its formation by percolation processes.\\

The two-fold exploration of Minkowski tensors, on the one hand in terms of measure theory and on the other in terms of surface integrals, is a genuine scientific achievement of this article and of significant potential for future research. The mathematical disciplines of integral geometry \cite{SchneiderWeil:2008} and stochastic geometry \cite{StoyanMeckeKendall:1995} are rich but (for non-experts) murky fishing grounds for research in random disordered systems. Many of the theorems relating to Minkowski functionals may well turn out to have physical counterparts or applications, following the examples of Hadwiger's theorem (that has been explored to describe shape dependence of the thermodynamics of confined fluids \cite{KoenigRothMecke:2004}) and the kinematic formula (needed for the development of density functionals for structured liquids \cite{Mecke:2000}). There are several current trends in integral and stochastic geometry with potential for physical applications. For example, the so-called {\em mixed functionals} are generalisations of the Minkowski functionals to functionals of two (or more) bodies, obtained by integration over two-body support measures, see chapter 6.4 in \cite{SchneiderWeil:2008}. The mixed measures are of likely benefit for the development of density functionals for fluids of aspherical particles without adjustable parameters. Current density functionals for non-spherical particles contain the ill-determined so-called $\zeta$-factor, which results from only retaining the first term of an expansion \cite{HansenGoosMecke:2010}. For aspherical particles, this expansion is needed as it leads to a factorisation into measures centred on a single body of the double integration over the translational and orientational degrees of freedom of the relative position of {\em two} fluid particles. Future work on advanced methods to evaluate mixed measures may help avoid the need for this expansion, and lead to a deeper understanding of the geometric principles that govern the thermodynamic properties of aspherical fluids. By making the concise but, for many, unfamiliar notation of integral geometry accessible to the reader, this article may facilitate future applications of integral geometry in physics and material science. \\

How do you measure the ``shape'' of an amorphous structure? This question expresses the need to identify a small number of morphological measures or structure metrics that, when evaluated for a spatial structure, capture its essential features in a few numbers. Which features are essential evidently depends on the physical property of interest, so a universal, generally valid answer to this question of the best structure metrics cannot be given. Nevertheless, some morphological properties recurrently appear as relevant to many physical processes. Some of these, such as densities or occupied volumes, areas of interfacial surfaces or spatial connectedness, turn out to be closely related to the scalar Minkowski functionals. The study of such quantities within an encompassing mathematical framework (here the theory of valuations in integral geometry) often gives new insight for the morphological analysis of the physical system; for example, the discussion of the relationship between $d$-dimensional percolation and the Euler index $\chi$ in the context of Minkowski functionals and integral geometry \cite{MeckeSeyfried:2002,NeherMeckeWagner:2007} has contributed to the increasingly wide-spread use of $\chi$ as a measure of spatial connectedness in the physical sciences. The family of {\em tensorial} Minkowski functionals, at the heart of this article, have several properties that make them suitable generic shape measures to capture morphological aspects related to anisotropy, elongation and orientation: they are a natural generalisation of the concepts of volume and surface area to tensor-valued quantities, benefit from a definition widely applicable to different types of geometry and from significant mathematical theory, and are closely related to two already widely used tensors, namely the tensor of inertia and the interface tensor. We anticipate that these tensorial shape measures will be identified as the relevant morphological descriptors in a growing variety of physical systems. The analytical and algorithmic methods derived in this article will provide the tools for wide-spread use of Minkowski tensors analyses in physics, material science, biological imaging and other disciplines. \\

\section*{Supplementary online material and software}
Software to compute the Minkowski tensors is made available as supplementary material to this article. The software and possible future extensions are also available through the website {\tt www.theorie1.physik.fau.de/karambola}.

\section*{Acknowledgement}

We are grateful to T.~Aste, G.~Delanay, M.~Saadatfar, T.~Senden and
M.~Schr\"oter for the bead pack data, to J.~Liu and D.A.~Weitz for the actin
biopolymer data, and M.~Spanner for Monte Carlo data of equilibrium hard sphere systems. We thank C.~Marlow for permission to
reproduce his painting ``White Spirits'', A.~Boyde for the
trabecular bone image, M.~Saadatfar for the
metal foam image and A.~B\"oker and V.~Oszowka for the copolymer film image in Fig.~\ref{fig:examples-of-anisotropic-3D-structure}.
We thank J.~H\"{o}rrmann and M.~Klatt for their critical comments on the manuscript, and M.~Hoffmann for help with Voronoi computations. GEST, DH and KM acknowledge support by the German
research foundation (DFG) through the research group `Geometry and Physics of
Spatial Random Systems' under grants SCHR1148/3-1, ME1361/12-1 and HU1874/2-1.

\section*{Appendix: Specific examples\label{app:examples}}

For some simple shapes the rank-2 MT can be calculated
analytically by
using explicit surface parametrisation and expressions for surface normals and
principal curvatures. Specifically for a sphere of radius $R$ centred
at the origin, one obtains 
\begin{equation}
W_0=\frac{4\pi }{3}R^3,\quad W_0^{2,0}=\frac{4\pi }{15}R^5 Q
\end{equation}
and, for $\nu=1,2,3$ and $r+s=2$, 
\begin{equation}
W_\nu=\frac{4\pi }{3}R^{3-\nu}, \quad  W_\nu^{r,s}=
\frac{4\pi }{9}R^{3-\nu+r}Q
\label{eq:tensors-sphere}
\end{equation}
with the unit tensor $Q$. 

For a convex polytope $P$, we write $\mathcal{F}_\mu(P)$ for the set of
$\mu$-dimensional 
faces of $P$, $\mu=0,1,2$, that is, $\mathcal{F}_0(P)$ is the set of 
vertices, $\mathcal{F}_1(P)$ is the set of 
(\emph{non-oriented}) edges, and $\mathcal{F}_2(P)$ is the set of faces. If
$F\in \mathcal{F}_\mu(P)$, 
then we denote by $\mathbf{n}(P,F)$ the set of exterior unit normal vectors 
of $P$ at $F$, which is a $(2-\mu)$-dimensional subset of the unit sphere
$\mathbbm{S}^2$. 
Then we obtain, as a 
special case of general formulae in section
\ref{subsec:def-of-MT-fundam-measure},
\begin{equation}\label{Tensorforpolytopes}
W_\nu^{r,s}(P)=\frac{1}{3}\sum_{F\in\mathcal{F}_{3-\nu}(P)}\int_F\mathbf{x}^r 
\mathcal{H}^{3-\nu}(\d\mathbf{x}) \int_{\mathbf{n}(P,F)}
\mathbf{n}^s \mathcal{H}^{\nu-1}(\d\mathbf{n})
\end{equation}
with $\nu=1,2,3$. For the notation see the main text at eq.~(\ref{eq:LocalMeasuresPolytope}).

For a cuboidal box of dimensions $a_x\times a_y\times a_z$ aligned with the
coordinate axes and centred at the origin eq.~(\ref{Tensorforpolytopes}) yields
$W_0=a_x a_y a_z$, $W_1=\frac{2}{3}(a_x a_y+a_y a_z+a_z a_x)$,
$W_2=\frac{\pi}{3}(a_x+a_y+a_z)$, $W_3=\frac{4\pi}{3}$, $W_\nu^{1,0}=0$ and all
MT 
of rank two are diagonal matrices with the following entries

\begin{eqnarray}
(W_0^{2,0})_{ii}=\frac{1}{12}a_i^3 a_j a_k,\\
(W_3^{2,0})_{ii}=\frac{\pi}{3}a_i^2,\\
(W_1^{2,0})_{ii}=\frac{a_i^3(a_j+a_k)}{18}+\frac{a_i^2a_ja_k}{6},\\
(W_1^{0,2})_{ii}=\frac{2 a_j a_k}{3}, \\
(W_2^{2,0})_{ii}=\frac{\pi}{36}\left(a_i^3+3 a_i^2(a_j+a_k)\right), 
(W_2^{0,2})_{ii}=\frac{\pi}{6} (a_j+a_k) 
\end{eqnarray}
where  $\{i,j,k\}=\{x,y,z\}$ and permutations thereof.

A torus centred at the origin with major radius $R_1$ 
and minor radius $R_2\le R_1$ can be parametrised by
$\mathbf{x}(u,v)=\{\cos(u)\left(R_1+R_2\cos(v)\right),
\sin(u)\left(R_1+R_2\cos(v)\right),R_2\sin(v)\}$ with
$\alpha,\beta\in[0,2\pi)$.  The scalar functionals are explicitly given by
$W_0=2\pi^2R_1 R_2^2$, $W_1=\frac{4\pi}{3}R_1 R_2$, $W_2=\frac{2\pi^2}{3}R_1$
and $W_3=0$ and the vectorial measures are $W_\nu^{1,0}=0$. The tensors of rank
two are diagonal with degenerate eigenvalues
$(W_\nu^{r,s})_{xx}=(W_\nu^{r,s})_{yy}$; the entries are given by
\begin{center}
\begin{tabular}{ll}
$(W_0^{2,0})_{xx} =  \frac{\pi^2}{4}R_1 R_2^2(4R_1^2+3R_2^2)$, &
$(W_0^{2,0})_{zz} =  \frac{\pi^2}{2}R_1 R_2^4$ \\[0.2cm]
$(W_1^{2,0})_{xx} =  \frac{\pi^2}{3}R_1 R_2(2R_1^2+3R_2^2)$,   &
$(W_1^{2,0})_{zz} =  \frac{2\pi^2}{3}R_1 R_2^3$ \\[0.2cm]
$(W_2^{2,0})_{xx} =  \frac{\pi^2}{6}R_1(2R_1^2+5R_2^2)$,       &
$(W_2^{2,0})_{zz} =  \frac{\pi^2}{3}R_1 R_2^2$ \\[0.2cm]
$(W_3^{2,0})_{xx} =  \frac{2\pi^2}{3}R_1 R_2$,                 &
$(W_3^{2,0})_{zz} =  0 $ \\[0.2cm]
$(W_1^{0,2})_{xx} =  \frac{\pi^2}{3}R_1 R_2$,                  &
$(W_1^{0,2})_{zz} =  \frac{2\pi^2}{3}R_1 R_2$ \\[0.2cm]
$(W_2^{0,2})_{xx} =  \frac{\pi^2}{6}R_1$,                      &
$(W_2^{0,2})_{zz} =  \frac{\pi^2}{3}R_1.$\\[0.2cm]
\end{tabular}
\end{center}

\begin{figure}[t]
\hfill\includegraphics[width=0.7\textwidth]{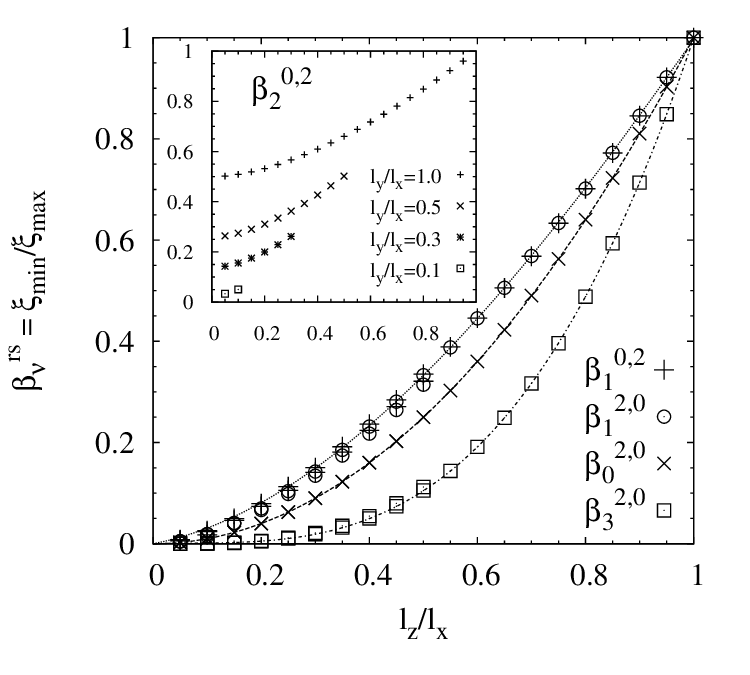}
\caption{Eigenvalue ratio of the smallest and largest eigenvalues $\xi_{\mathrm{min}}$
and $\xi_{\mathrm{max}}$ of the MT $W_\nu^{r,s}$ of an ellipsoid with
radii $l_x=1$ and $l_x=1\ge l_y \ge l_z$ as function of $r=l_z/l_x$. Each symbol
in the main plot represents data (hardly distinguishable) for three different
intermediate radii $l_y=0.1,0.5,0.9$ indicating that for these four tensors the
minimal to maximal eigenvalue ratio is approximately the same for all values of
the intermediate radius. The solid curves are fits to the data giving
$\beta_3^{2,0}\approx 1.210r^3-0.235r^2+0.024$, $\beta_0^{2,0}=r^2$,
$\beta_1^{2,0}\approx \beta_1^{0,2}\approx -0.366r^3+1.222r^2+0.139r$. The
insert shows the eigenvalue ratio of the tensor $W_2^{0,2}$ as function of
$l_z/l_x$. In contrast to the above four tensors, this ratio depends strongly on
the value of the intermediate radius $l_y$. In particular, for $l_z=0$ the
eigenvalue ratio only becomes zero if the intermediate radius is also $l_y=0$.
For the maximal $l_y=1$ the eigenvalue ratio converges to $0.5$ for
$l_z/l_x\rightarrow 0$. }
\label{fig:eig-val-ratio-ellipsoid}
\end{figure}

For an ellipsoid given by $(x/l_x)^2+(y/l_y)^2+(z/l_z)^2=1$ the surface
integrals all result in elliptic integrals and cannot be expressed in closed
form. However, the scalar MF $W_0$ is $W_0=\frac{4\pi}{3}l_xl_yl_z$. The MT $W_0^{2,0}$ is diagonal with
\begin{equation}
(W_0^{2,0})_{ii}=\frac{4\pi}{15}l_i^3l_jl_k
\end{equation}
where $\{i,j,k\}$ is $\{x,y,z\}$ and permutations thereof. The
integration of all other tensors is easily numerically obtained by use of the
ellipsoid parametrisation
$\mathbf{x}(u,v)=\{l_x\cos(u)\sin(v),l_y\sin(u)\sin(v),l_z\cos(v)\}$ which yields
explicit expressions for the metric tensor of the ellipsoidal surface, the
normal vector, and the mean and Gaussian curvatures. These are readily
integrated numerically. Fig.~\ref{fig:eig-val-ratio-ellipsoid} shows the minimal
to maximal eigenvalue ratio of the MT of rank two of ellipsoids
with $l_x=1$ and $1\ge l_y\ge l_z$.

\section*{References}
%\bibliographystyle{unsrt}
%
%\bibliography{./LITERATURE/literatur,TPMS}

\end{document}